\documentclass[]{aastex631}
\accepted{April 17, 2024}

\shorttitle{AASTeX v6.3.1 Sample article}
\shortauthors{Silva-Farfán et al.}
\graphicspath{{./}{figures/}}
\usepackage{cprotect}
\usepackage{natbib}

\defcitealias{forster2018delay}{F18}

\defcitealias{Martinezcorr}{M22}

\defcitealias{Subrayan}{S22}

\begin{document}

\title{Physical Properties of Type II Supernovae Inferred from ZTF and ATLAS Photometric Data}

\newcommand\MAS{Millennium Institute of Astrophysics (MAS), Nuncio Monse\~nor Sotero Sanz 100, Of. 104, Providencia, Santiago, Chile}
\newcommand\CMM{Center for Mathematical Modeling, Universidad de Chile, Beauchef 851, Santiago 8370456, Chile}
\newcommand\IDIA{Data and Artificial Intelligence Initiative (IDIA), Faculty of Physical and Mathematical Sciences, Universidad de Chile, Chile.}
\newcommand\DAS{Departamento de Astronom\'{\i}a, Universidad de Chile, Chile}

\author[0009-0000-2457-279X]{Javier Silva-Farfán}
\affiliation{\DAS}

\author[0000-0003-3459-2270]{Francisco F\"orster}
\affiliation{\IDIA}
\affiliation{\MAS}
\affiliation{\CMM}
\affiliation{\DAS}

\author[0000-0003-1169-1954]{Takashi J. Moriya}
\affiliation{National Astronomical Observatory of Japan, National Institutes of Natural Sciences, 2-21-1 Osawa, Mitaka, Tokyo 181-8588, Japan}
\affiliation{School of Physics and Astronomy, Monash University, Clayton, Victoria 3800, Australia}

\author[0000-0002-8606-6961]{L. Hern\'andez-Garc\'ia}
\affiliation{\MAS}
\affiliation{Instituto de F\'isica y Astronom\'ia, Facultad de Ciencias,Universidad de Valpara\'iso, Gran Breta\~na No. 1111, Playa Ancha, Valpara\'iso, Chile}

\author[0000-0002-8722-516X]{A. M. Mu\~noz Arancibia}
\affiliation{\MAS}
\affiliation{\CMM}

\author[0000-0003-0820-4692]{P. S\'anchez-S\'aez}
\affiliation{European Southern Observatory, Karl-Schwarzschild-Strasse 2, 85748 Garching bei München, Germany}
\affiliation{\MAS}

\author[0000-0003-0227-3451]{Joseph P. Anderson}
\affiliation{European Southern Observatory, Alonso de C\'ordova 3107, Casilla 19, Santiago, Chile}
\affiliation{\MAS}

\author[0000-0003-2858-9657]{John L. Tonry}
\affiliation{Institute for Astronomy at University of Hawaii}

\author{Alejandro Clocchiatti}
\affiliation{Instituto de Astrof\'isica, Facultad de F\'isica, Pontificia Universidad Cat\'olica de Chile, Av. Vicuña Mackenna 4860, Santiago, Chile}
\affiliation{\MAS}



\newcommand\nSNe{186}

\begin{abstract}


We report an analysis of a sample of 186 spectroscopically confirmed Type II supernova (SN) light curves (LCs) obtained from a combination of Zwicky Transient Facility (ZTF) and Asteroid Terrestrial-impact Last Alert System (ATLAS) observations. We implement a method to infer physical parameters from these LCs using hydrodynamic models that take into account the progenitor mass, the explosion energy, and the presence of circumstellar matter (CSM). The CSM is modelled via the mass loss rate, wind acceleration at the surface of the progenitor star with a $\beta$ velocity law, and the CSM radius. We also infer the time of explosion, attenuation (A$_V$), and the redshift for each SN. Our results favor low-mass progenitor stars (M$_{ZAMS}$\,$<$14\,$M_\odot$) with a dense CSM ($\dot{M}$ $>$ 10$^{-3}$ [M$_\odot$ yr$^{-1}$], a CSM radius of $\sim$ 10$^{15}$ cm, and $\beta$ $>$ 2). Additionally, we find that the redshift inferred from the supernova LCs is significantly more accurate than that inferred using the host galaxy photometric redshift, suggesting that this method could be used to infer more accurate host galaxy redshifts from large samples of SNe II in the LSST era. Lastly, we compare our results with similar works from the literature.

\end{abstract}


\keywords{Supernovae, Type II supernovae, Surveys,}


\section{Introduction} \label{sec:intro}


A supernova (SN) is an event that occurs when a star ends its life abruptly, typically in an explosion with an energy of $\sim$ 10$^{51}$ erg. SNe are classified by features in their spectra into two main classes: Type II SNe (hereafter SNe II) that show features of hydrogen in their spectra; and type I SNe that do not. Hydrogen in the spectra of SNe II arises from the progenitor's envelope \citep{minkowski, filippenko_spec}. The progenitors of SNe II are red supergiant (RSG) stars, with zero-age main sequence (ZAMS) mass $\gtrsim 8 M_{\odot}$ \citep{1996ASPC...98..220L}.

SNe II are the most common type of SNe in nature \citep{li2011nearby} and can help us understand the latest stages of stellar evolution for massive stars, e.g. the mass loss before explosion. To study these objects, we can directly identify the progenitor in archival images \cite[see][]{Smartt2009}, but this is a difficult task due to how faint these stars are compared to SNe and is only possible for the most nearby objects. 


The emergence of different surveys over the past years, such as the Palomar Transient Factory (PTF) \citep{PTF}, the Panoramic Survey Telescope and Rapid Response System (Pan-STARRS) \citep{Pan-STARRS}, the All Sky Automated Survey for SuperNovae (ASAS-SN) \citep{ASAS-SN}, the Zwicky Transient Facility (ZTF) \citep{ztf}, and the Asteroid Terrestrial-impact Last Alert System (ATLAS) \citep{atlas} have rapidly increased the number of discovered SNe. The upcoming Vera C. Rubin Observatory Legacy Survey of Space and Time \citep[LSST,][]{LSST}, is forecast to discover millions of SNe during its ten-year duration, thus increasing the number of SNe discovered as we have never seen before. However, since spectroscopic follow-up capabilities will be limited, the development of methods to study these objects using only photometric data is required. Therefore, the development of techniques that allow us to study large samples of objects and analyze their physical properties using exclusively light curve (LC) information will be necessary.

To prepare for the coming deluge of data from e.g. LSST, not only are techniques to analyze big data necessary, but also theoretical studies to understand the physics behind the objects. In the case of SNe II, the link between the progenitor and the SN is still not fully understood. Theoretical LC modeling of SNe II has shown that these objects must have an extremely large mass loss rate in order to explain the rise times of their LCs \citep{Morozova2018}. Also, there is evidence that the progenitors of SNe II have a dense circumstellar matter (CSM) near the surface of the star \citep{Khazov2016, yaron2017confined}. The mechanism through which SN II progenitors produce this dense CSM before exploding is not clear, although there are a couple of scenarios that could explain it: 1) pre-explosion outburst: The energy deposited by waves generated from late-stage nuclear burning is proposed to drive the ejection of material from the star's outer layers producing a dense CSM in the final years before the explosion \cite[see, e.g.][]{Fuller2017, Morozova2020}; 2) enhanced density in the vicinity of the progenitor due to wind acceleration: \cite{Moriya2017, Moriya2018} proposed a wind accelerated scenario, where the wind follows a $\beta$ velocity law \citep{1975ApJ...195..157C} in the last $\sim$ 100 years before the explosion, the slowest wind closest to the surface of the star can achieve much higher densities, similar to those in the outburst model without requiring extremely high mass loss rates. For a more detailed discussion see \citet{2022MNRAS.517.1483D}.
 


\citet[][hereafter \citetalias{forster2018delay}]{forster2018} introduced a method to infer physical parameters using hydrodynamical models in order to classify SN from the High cadence Transient Survey (HiTS) using their LCs.

In this work, we will use the method from \citetalias{forster2018delay} using hydrodynamical models from \citet[][hereafter M18]{Moriya2018}. We would like to clarify that when we refer to SNe II during this work we refer to type IIP and type IIL events, as the hydrodynamical models used in this work were developed for the progenitors of these types of SNe. We do not incorporate other subtypes such as type IIn, type IIb, or type IIc in this work. We start by adapting and optimizing the method, and then we infer the physical parameters of a sample of spectroscopically confirmed SNe II using a Bayesian approach on publicly available forced photometry data from ZTF and ATLAS. We study the distribution and correlations of the inferred physical parameters in the sample and compare our results with independent studies. We also compare the inferred redshifts with those available from spectroscopic and photometric host information to validate our results. Finally, we  discuss the limitations of our method and its applicability in the LSST era.


This paper is organized as follows: in Section \ref{sec:sample} we present how we select the sample of spectroscopically confirmed SNe II. In Section \ref{sec:method} we explain our method to infer physical parameters. In Section \ref{sec:Results} we present the results of our method applied to the sample, the distribution of the inferred parameters, and a comparison between the redshift inferred using our method and the redshift of the host galaxy. In Section \ref{sec:Analysis} we discuss the results obtained in Section \ref{sec:Results}, we compare our results with other works, discuss the possibility of our method being used as a distance indicator, the limitations of our method and the implementation on LSST data. Lastly, in Section \ref{sec:Conclusion} we present the conclusions of this work and discuss future work.




\section{Sample} \label{sec:sample}

The SNe II sample used in this work is defined using both spectroscopic and photometric criteria. It is important to note that a spectroscopic confirmation can be sometimes incorrect, since the classification accuracy depends on the instrument resolution and the phase when the spectra was taken. First, we employ The Automatic Learning for the Rapid Classification of Events (ALeRCE) broker light curve classifier \citep{alerce_general, ALeRCE_LC} to select all the objects classified as SNII with a probability higher than 0.3. At the time when this was sample was defined (April 2021), 452 SNe candidates were selected. This process allowed us to search for SNe that have a minimum number of detections (at least 6 detections) in a given ZTF band and that are photometrically consistent with being Type II SNe.

Then, we searched for SNe in our sample that were spectroscopically classified as SNe II. This step was done by crossmatching with the Transient Name Server (TNS\footnote{\url{https://www.wis-tns.org/}}) database and discarding the objects that were not classified as SNe II. We found 252 confirmed SNe II in the 452 candidates sample. Of the 252 confirmed SNe, we discarded those that have a gap larger than 10 days between the last non-detection (where the object's brightness variation did not exceeded the signal-to-noise threshold) and the first detection (where the object's measured  brightness variation surpassed the signal-to-noise threshold). This allows for better constraints on the explosion time. The final sample is composed of 186 SNe II.


To acquire the necessary data, we utilized the forced photometry services provided by both ZTF \citep[see][]{ztf_forcedphot} and ATLAS \citep[][]{atlas, atlas_transient}\footnote{\url{https://fallingstar-data.com/forcedphot/}}. By supplying the mean coordinates from the ZTF objects (obtained through the ALeRCE database) and specifying the desired time range, data was downloaded covering 50 days preceding the first ZTF alert up to the date of the data request.

\subsection{Light curve cleaning}

\subsubsection{ZTF forced photometry}
Now, we describe how we process ZTF forced photometry data to discard outliers and large error data points. The procedure presented in this section is similar to the one from \cite{2023MNRAS.tmp..753H}. The following criteria were found to satisfactorily clean the data. We start by following the guidelines from \cite{ztf_forcedphot}, filtering by the per-epoch processing status code \verb|procstatus| variable. We keep any observation with a \verb|procstatus| value equal to 0 (successful execution), or 56 (one or more epochs have photometric measurements that may be impacted by bad pixels), or 57 (one or more epochs had no reference image catalog source falling within 5 arcsecs). Any observation where the CCD-quadrant-based image quality \verb|infobits| variable was not equal to 0 was removed. 

Additionally, we perform quality cuts following the ZTF Science Data System (ZSDS) Advisories and Cautionary Notes, from the ZTF DR5 Documentation, section 2.4, flagging as a bad data point any data point that satisfies the following conditions for the zeropoint magnitude (ZP):

\begin{minipage}[t]{0.45\linewidth}
For g filter
    \begin{itemize}
        \item ZP $>$ 26.7 - 0.2 $\times$ airmass OR
        \item ZP rms $>$ 0.06 OR
        \item ZP $<$ threshold[ccd] - 0.2 $\times$ airmass
    \end{itemize}
\end{minipage}\hfill
\begin{minipage}[t]{0.45\linewidth}
For r filter
    \begin{itemize}
        \item ZP $>$ 26.65 - 0.15 $\times$ airmass OR
        \item ZP rms $>$ 0.05 OR
        \item ZP $<$ threshold[ccd] - 0.15 $\times$ airmass
    \end{itemize}
\end{minipage}

\vspace{1mm}

\noindent where threshold[ccd] is  the zero point thresholds used to identify bad quality images and varies depending on the CCD used for the observation and the filter, and ranges from  25.6712 to 25.9225 for the g-filter and from 25.6199 to 25.9759 for the r-filter. Finally, for those cases when the SN was already in the difference image, the early and/or late section of the light curve exhibited a negative difference flux. Consequently, a baseline correction, as outlined in \citet{ztf_forcedphot}, was applied when possible (when observations before the explosion are available to define the new baseline).

\subsubsection{ATLAS forced photometry}

The ATLAS reduction pipeline from the alerts system has a custom built point-spread-function (PSF) fitting routine that runs on the difference images to produce flux measurements of all sources that are detected at 5$\sigma$ or more above the background noise. This routine is called tphot and is based on the algorithms discussed in \cite{Tonry2011, Sonnett2013}. The ATLAS forced photometry service started in 2021, providing public access to photometric measurements over the full history of ATLAS survey, thus allowing us to access data below 5$\sigma$ of background noise.

At the time of this work and unlike the case of ZTF, there were no guidelines available to clean ATLAS forced photometry LCs. Consequently, we explored the available metrics to discard contaminated observations with large error bars. We keep observations that meet all the following conditions:
\begin{itemize}
    \item 0.5 $<$ $\chi^2$/N $<$ 3
    \item flux $>$ -100 [$\mu$Jy]
    \item sky magnitude o filter $>$ 18, sky magnitude c filter $>$ 18.5
    \item flux error $<$ 40 [$\mu$Jy]
\end{itemize}
where $\chi^2$/N is the reduced $\chi^2$ of the PSF fit, sky magnitude is the sky magnitude in 1 square arcsec, flux is the forced photometry PSF flux of the different image measured in microJanskys, and flux error is the reported error for the flux.

\begin{figure}[h!]
	\centering
	\includegraphics[scale=.45]{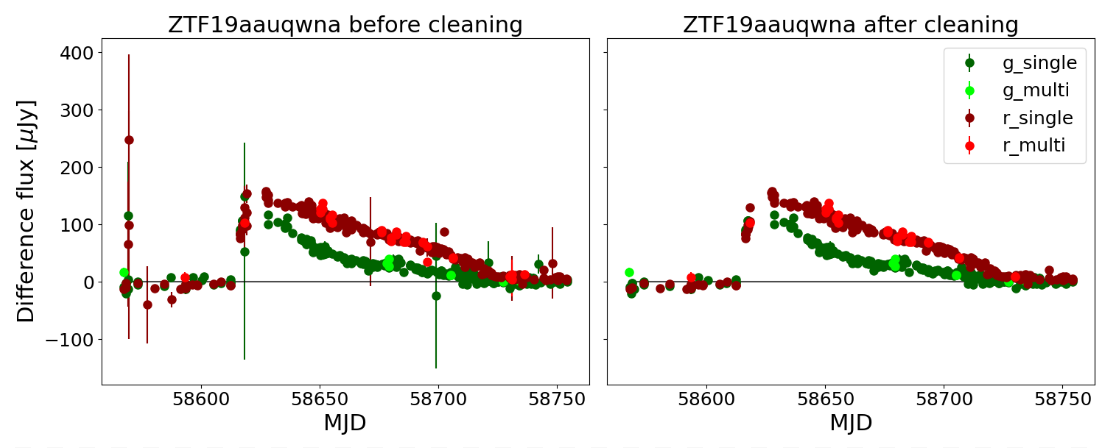}
	\caption{SN2019fem/ZTF19aauqwna ZTF LC before and after cleaning are shown in the left and right panel, respectively. The single and multi suffixes for the g and r bands correspond to the type of coating on the ZTF CCDs. These different coatings result in differences in the transmission curves, as explained in Section \ref{sec:lightcurves}}
	\label{ZTF_cleaning}
\end{figure}

\begin{figure}[h!]
	\centering
	\includegraphics[scale=.45]{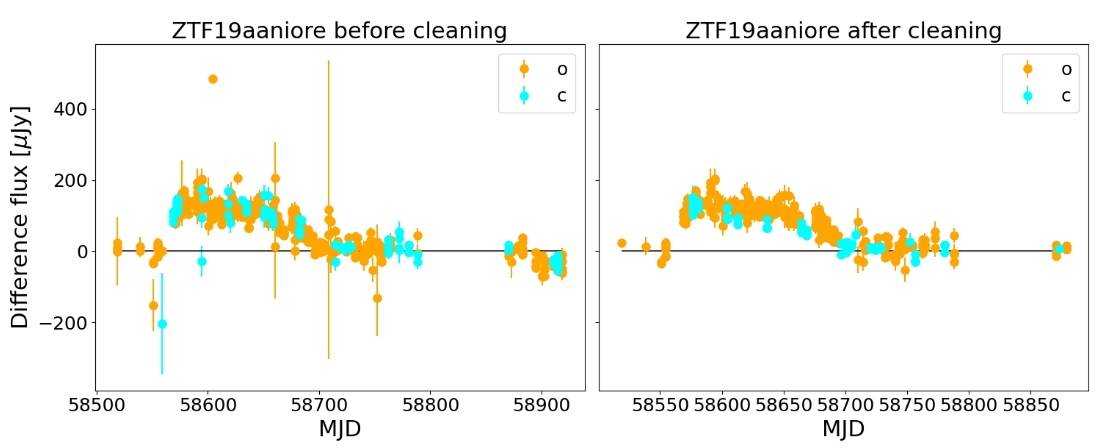}
	\caption{SN2019ceg/ZTF19aaniore ATLAS LC, before and after cleaning are shown in the left and right panel respectively.}
	\label{ATLAS_cleaning}
\end{figure}

In Figures \ref{ZTF_cleaning} and \ref{ATLAS_cleaning} we show examples of LCs before and after the cleaning procedure is applied, respectively. 

\section{Methodology} \label{sec:method}

Our method is a modification of the work from \citetalias{forster2018delay}, adapted to be used with data from ZTF and ATLAS and optimized to be $\sim$ 6 times faster using the python package Numba \citep{numba}. We use a Bayesian approach to infer physical parameters from a SN LC given models for the explosion and the telescope. The method starts with time series of spectra and generates synthetic LCs given the transmission curves of the telescope. This is used to build a grid of synthetic LCs that are interpolated in order to evaluate likelihoods at arbitrary parameter combinations.


\subsection{LC models}
M18 introduced synthetic LC models of Type~II SNe that are computed by one-dimensional multi-frequency radiation hydrodynamics code \texttt{STELLA} \citep{Blinnikov1998,Blinnikov2000,Blinnikov2006}. It evaluates the evolution of spectral energy distributions (SEDs) in each timestep and we can obtain synthetic LCs of any given filters by convolving filter functions with the synthetic SED evolution. Given a progenitor model, we assume a mass cut of 1.4~$\mathrm{M_\odot}$ and put an assumed $^{56}$Ni at the central region by hand. Then we initiate an explosion by putting a thermal energy just above the mass cut.

M18 takes the effects of CSM into account to compute LCs. The CSM structure is from the red supergiant (RSG) winds and is constructed
by adopting a $\beta$ velocity law, described by the following equation:
\begin{equation} \label{beta_vel}
v_{wind}(r) = v_{0} + (v_{\infty}- v_{0}) \left(1 - \frac{R_{0}}{r}\right)^{\beta}, 
\end{equation}
where $v_0$ is the ejection velocity (chosen to smoothly connect the density from the surface of the progenitor and the wind), $v_\infty$ is the terminal wind velocity (set to 10 km s$^{-1}$), $R_0$ is the ejection radius (set at the stellar surface), and $\beta$ is the velocity law index. The parameters of this model are a) ZAMS mass of the progenitor, b) energy of the explosion, c) the mass loss rate ($\dot{M}$), d) the CSM radius, and e) $\beta$ from eq. \ref{beta_vel}. We used a grid of 1686 models from M18 with parameters listed in Table~\ref{tabparams}. 
One limitation of these models is that they are only available for progenitor masses starting from 12 M$\odot$, whereas the progenitors of SNe II can have masses as low as 8 M$\odot$. This limitation arises from the fact that the models from M18 focused on the early part of the LC, where the mass did not affected the properties of the early LC, therefore a wider range was not necessary. In this work, we have not included or expanded the models, as we will discuss later, due to computational limitations. Also, these models have a fixed value for $^{56}$Ni mass of 0.1 M$_\odot$ and do not have photospheric velocity information unlike \citet[][hereafter \citetalias{Martinezcorr}]{Martinezcorr}. It is important to highlight the later because \citetalias{Martinezcorr} showed that ignoring the photospheric velocity can affect the infered values of energy and mass. Since we are attempting to understand the science that could be done with LSST data, where most SNe will lack spectroscopic information, we have decided not to use photospheric velocity information. However, in Section \ref{sec:comparisonM22S22} we will compare our results with those of \citetalias{Martinezcorr}, to gain insight into the potential impact of this effect.

\begin{deluxetable}{cccccccc}

\tablecaption{RSG progenitors parameters}
\tablenum{1}
\label{tabparams}

\tablehead{\colhead{Mass} & \colhead{Energy} & \colhead{Mass loss rate} & \colhead{CSM radius} & \colhead{$\beta$} & \colhead{} & \colhead{} & \colhead{} \\ 
\colhead{(M$_\odot$)} & \colhead{(foe)} & \colhead{(M$_\odot$ / year)} & \colhead{(10$^{15}$ cm)} & \colhead{} & \colhead{} & \colhead{} & \colhead{} } 

\startdata
12 & 0.5 & 0 & 0 & 0 &  \\
14 & 1 & 10$^{-5}$ & 0.1 & 1 &  \\
16 & 1.5 & 3  $\times$  10$^{-5}$ & 0.3 & 1.75 &  \\
 & 2 & 10$^{-4}$ & 0.5 & 2.5 &  \\
 & & 3  $\times$  10$^{-4}$ & 1 & 3.75 &  \\
 & & 10$^{-3}$ & & 5   \\
 & & 3 $\times$ 10$^{-3}$ &  \\
 & & 10$^{-2}$ & & \\
\enddata

\end{deluxetable}

\subsection{Synthetic Light curves} \label{sec:lightcurves}

To produce synthetic LCs for any redshift, attenuation, and explosion time from hydrodynamic models time series of spectra we assumed a standard $\Lambda$CDM model with $H_0$ = 70 km Mpc$^{-1}$ s$^{-1}$, and $\Omega_0$ = 0.3 and then the spectra time series are redshifted and attenuated assuming a Cardelli law with $R_V$ = 3.1 for dust attenuation. The spectra are integrated over the filter bandwidths from ATLAS and ZTF to generate synthetic LCs. In our work, we use the time series spectra generated from M18, which span from 1 to 47000 Angstrom. Our synthetic LCs are produced through a combination of the parameters from Table \ref{tabparams}, including roughly three times more models than those used \citetalias{forster2018delay} because of the inclussions of different values for the CSM radius.

\begin{deluxetable*}{cccc}


\tablecaption{Filters description}
\tablenum{2}
\label{tabfilters}

 \tablehead{
 \colhead{Filter} & \colhead{$\lambda_{mean}$ [nm]} & \colhead{W$_{eff}$} [nm]} 

 \startdata 
ZTF g & 483.5 & 117.7\\
ZTF r & 646.3 & 141.7\\
 ATLAS o & 686.6 & 236.8\\
 ATLAS c & 540.8 & 214.4
 \enddata
\tablecomments{$\lambda_{mean}$ and W$_{eff}$ are the mean wavelength and effective width respectively as defined in \citet[][]{2012ivoa.rept.1015R}}
\end{deluxetable*}

In the case of ATLAS, we generate synthetic LCs for the c filter and for the o filter. In the case of ZTF, we use two filters, ZTF g and ZTF r. Details about these filters are listed in Table \ref{tabfilters}. It is worth noting that CCDs in ZTF have two kinds of coating \cite[see][]{ztf_instrumentation}. These coatings influence the quantum efficiency and, consequently, the transmission curve. As a result, we generate four effective bandpasses from the original ZTF g and ZTF r filters. These new bandpasses are labeled as g single, g multi, r single, and r multi, where single and multi refer to CCDs that have a single or double layer coating, respectively.

\begin{deluxetable*}{ccccc}


\tablecaption{Pre-compute light curves parameters}
\tablenum{3}
\label{tabLcsResolution}

 \tablehead{
 \colhead{Parameter} & \colhead{Min value} & \colhead{Max value} & \colhead{Steps}} 

 \startdata 
Time [days] & 10$^{-3}$ & 1000 & 100\\
z & 10$^{-3}$ & 1 & 30\\
A$_{V}$ [mag] & 10$^{-4}$ & 10 & 10
 \enddata
\tablecomments{All parameters are evenly spaced on log scale}
\end{deluxetable*}

We pre-compute the LCs for all the bands and all available models in a logarithmically spaced Time (since explosion), A$_{V}$, and z arrays summarized on Table \ref{tabLcsResolution}, thus producing a total of 505,800 synthetic LCs as a combination of the model physical parameters, attenuation, and redshift. These synthetic LCs are then interpolated to explore the whole parameter space. To interpolate between models we used the model interpolation introduced in \citetalias{forster2018delay} (see Appendix~\ref{model_interpolation} for more details).



\begin{deluxetable*}{ccc}


\tablecaption{Prior distribution}
\tablenum{4}
\label{tabprior}

 \tablehead{
 \colhead{Parameter} & \colhead{prior distribution} & \colhead{Units}}

 \startdata 
 scale & N(1, 0.01) & \\
t$_{exp}$ & N(t$_{exp0}$, 4) & days\\
 ln z & N(ln 0.18, 2), z $\in$ (10$^{-3}$, 1) & \\
 ln A$_V$ & N(ln 0.05, 2), A$_V$ $\in$ (10$^{-4}$, 10) & ln mag.\\
 mass & N(14, 3), mass $\in$ (12, 16) & M$_\odot$\\
 energy & N(1, 1), energy $\in$ (0.5, 2) & B\\
 log$_{10}$ $\dot{M}$ & U(-8, -2), log$_{10}$ $\dot{M}$ $in$ (-8, -2) & log$_{10}$ M$_\odot$ yr$^{-1}$\\
 r$_{CSM}$ & N(0.5, 1), r$_{CSM}$ $\in$ (0.1, 1) & 10$^{15}$cm\\
 $\beta$ & N(3, 2), $\beta$ $\in$ (0, 5) & 
 \enddata

\tablecomments{N ($\mu$, $\sigma$) is a Gaussian distribution with mean $\mu$ and standard deviation $\sigma$, and U (a, b) is a uniform distribution between a and b. The prior probabilities are zero outside the intervals indicated.}


\end{deluxetable*}

\section{Results}  \label{sec:Results}


\begin{figure}[h!]
    \centering
	\includegraphics[scale=.34]{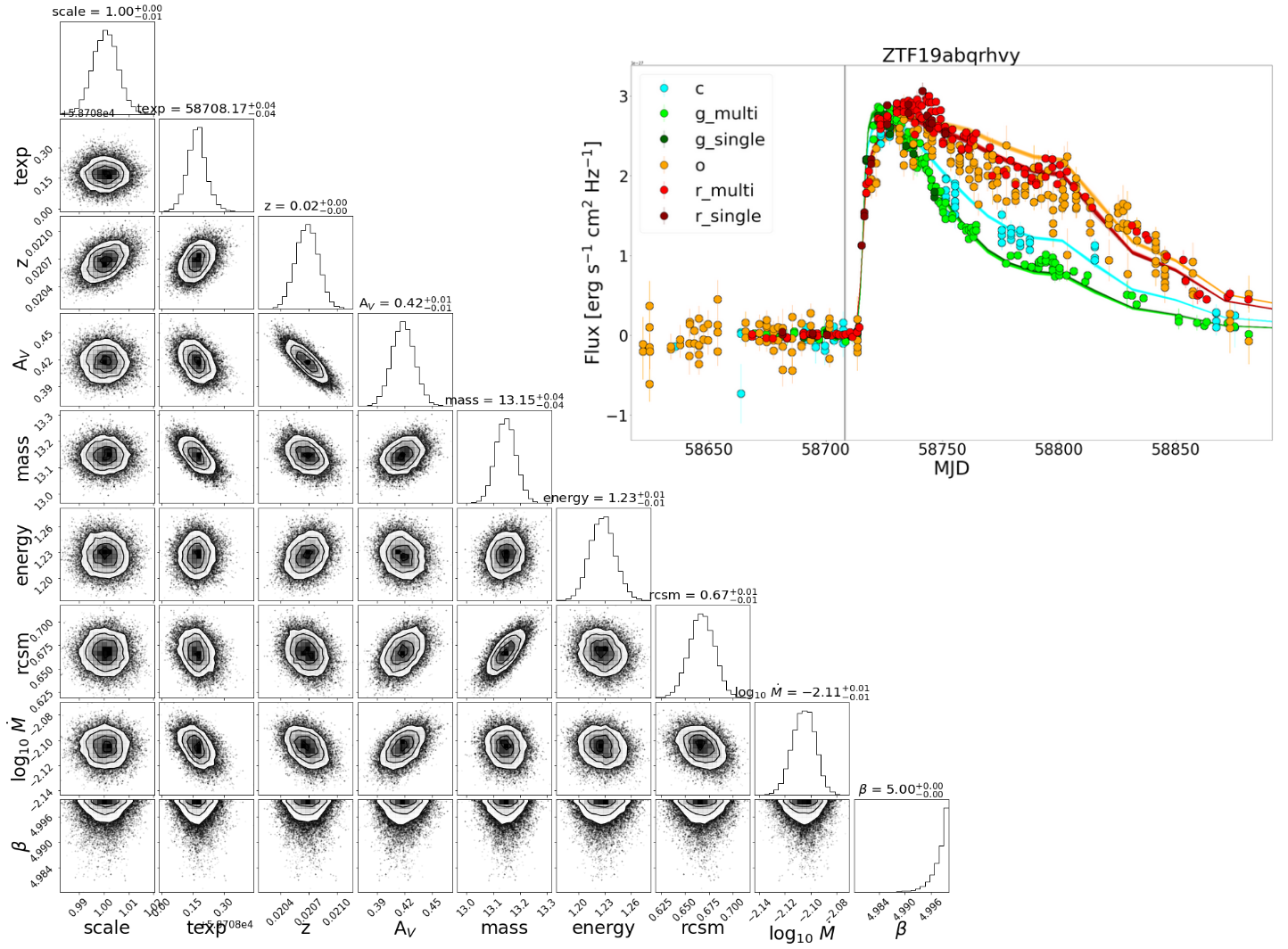}
	\caption{Bottom left corner: Corner plot of the posterior distribution of physical parameters obtained using our method for SN 2019odf/ZTF19abqrhvy. Top right corner: Observations of ZTF19abqrhvy (dots) and 100 models randomly sampled from the posterior LCs (continuous lines). Explosion times are indicated as grey vertical lines.}
	\label{posterior_example}
\end{figure}

Using our method with the priors from Table \ref{tabprior}, where scale is a parameter to allow for errors in absolute calibrations, t$_{exp}$ is the explosion time, t$_{exp0}$ is an estimation of the explosion that we define five days before the first detection, $\dot{M}$ is the mass loss rate, r$_{CSM}$ is the radius of the CSM, and $\beta$ is the exponent from the $\beta$ velocity law, we are able to obtain the posterior distribution of physical parameters for a sample of 186 SNe. We present an example corner plot showing the projected posterior distribution for SN 2019odf/ZTF19abqrhvy in Figure~\ref{posterior_example}, alongside the LC of the object with 100 random LCs sampled from the posterior distribution (thin continuous lines). This visual representation helps to evaluate how well the models match the data in a qualitative manner. Also, the explosion time for a given LC is plotted as a vertical grey line. In this case, the inferred parameters follow an apparently multivariate unimodal distribution, except for the marginal distribution of $\beta$ that reaches its maximum at the model limit.


For every parameter, we take the median from the marginalized posterior distribution as a representative value and the 5 and 95 percentiles as the lower and upper limits, respectively. We choose these values instead of the mean and standard deviation because not all posterior distributions are necessarily Gaussian and some posterior distributions are bimodal or multimodal. Thus, the median and percentiles provide a more robust and conservative description of their distribution.  The statistics of the marginalized posteriors for every object in the sample is shown in Table~\ref{tabSNe_params}.


\section{Analysis} \label{sec:Analysis}

\subsection{LC inferred redshift}

Following our methodology, the redshift can be left variable or fixed assuming its known value. We first leave it variable in order to validate our inference method with independently derived values of the redshift.
Different surveys have measured the redshift of galaxies and we can use information from their associated catalogs to compare the redshifts of SN host galaxies with our inferred values of redshift (the median of the marginalized posterior distribution, hereafter z$_{\rm LC}$). We use data from Sloan Digital Sky Survey data release 16 \citep{SDSSdr16}, \citet{https://doi.org/10.26132/ned1}, and SIMBAD \citep{SIMBAD} to obtain the redshifts of host galaxies.

\begin{figure}[ht!]
	\centering
	\includegraphics[scale=.14]{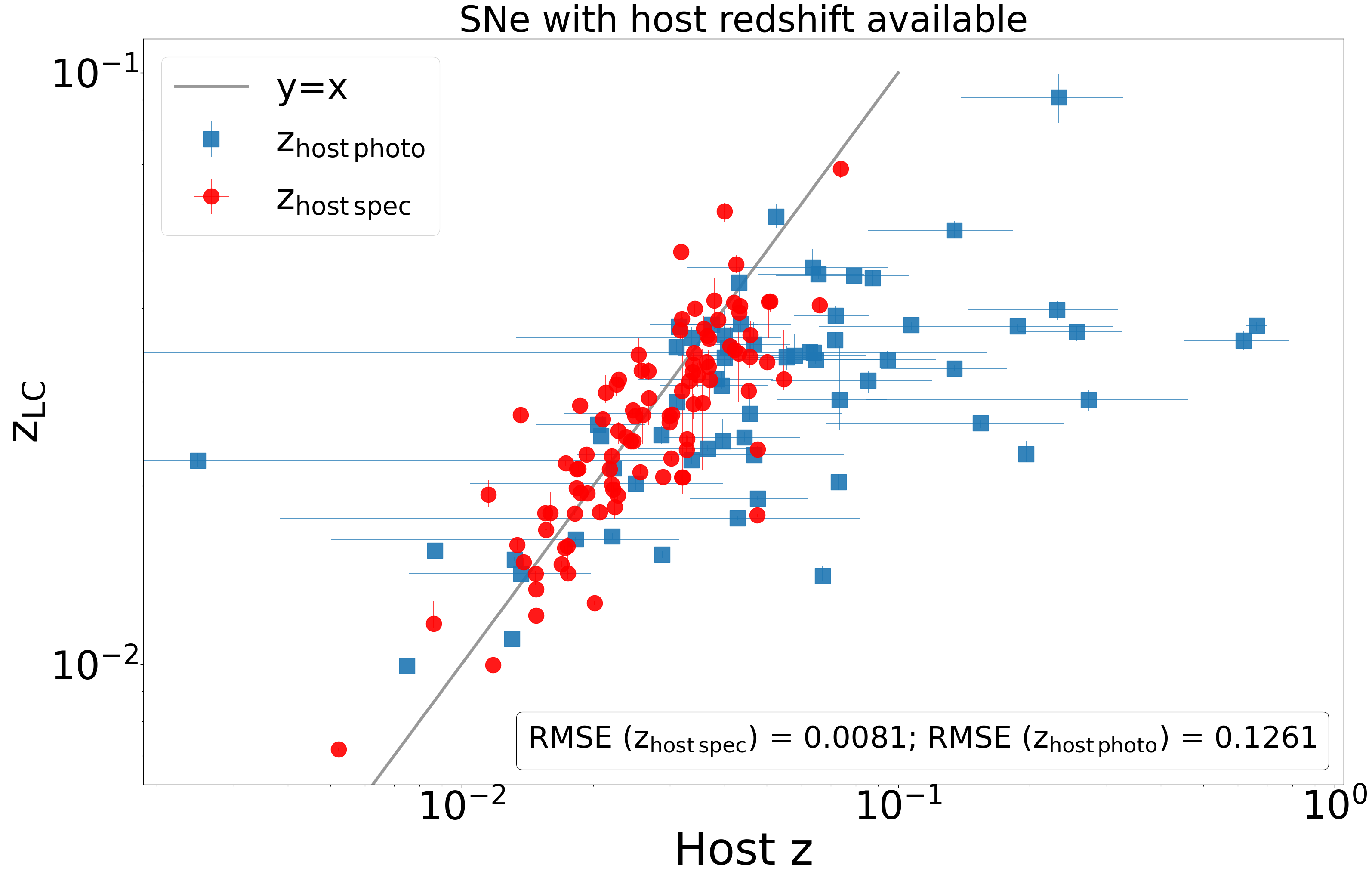}
	\caption{Relation between z$_{\rm LC}$ and z$_{\rm host \, spec}$, the error is reported as error bars when available. Red circles are host galaxies whose best redshift available was spectroscopically measured, while blue squares are host galaxies whose best redshift available was z$_{\rm host \, photo}$. RMSE for z$_{\rm LC}$ compared to z$_{\rm host \, spec}$ and z$_{\rm host \, photo}$ is reported at the bottom right corner.}
	\label{zVSz}
\end{figure}

To identify SNe with galaxy redshifts, we visually associate a host galaxy to every SNe in our sample (see Appendix~\ref{Appendix_hostassoc}) and obtain the best available redshift measurement (spectroscopic if available, photometric otherwise). Host spectroscopic redshifts (z$_{\rm host \, spec}$), red circles in Figure \ref{zVSz}, were obtained for 98 SNe from our sample; 64 SNe only had photometric redshifts (z$_{\rm host \, photo}$) available, blue squares in Figure \ref{zVSz}); and the remaining 24 SNe could not be associated with a host or the host did not have a z$_{\rm host \, spec}$ or z$_{\rm host \, photo}$ available. We found that z$_{\rm LC}$ is comparable with the z$_{\rm host \, spec}$ as seen in Figure \ref{zVSz}, where the root-mean square error (RMSE) for only spectroscopic redshift is RMSE = 0.0081, while comparing z$_{\rm LC}$ to z$_{\rm host \, photo}$ we obtain an RMSE = 0.1261.

When comparing the relation between z$_{\rm LC}$ and z$_{\rm host \, photo}$ in Figure \ref{zVSz}, we see a large scatter around the identity line. To test if this poor correlation is related to our inferred method or due to the low accuracy of z$_{\rm host \, photo}$ we compare both inferred redshifts, z$_{\rm host \, phot}$ and z$_{\rm LC}$, to z$_{\rm host \, spec}$. To do this we look for the cases in the sample of 98 SNe whose host has a z$_{\rm host \, spec}$ (red circles in Figure \ref{zVSz}) and a z$_{\rm host \, photo}$ for that host (72 out of the 98 SNe).

\begin{figure}[ht!]
	\centering
	\includegraphics[scale=.14]{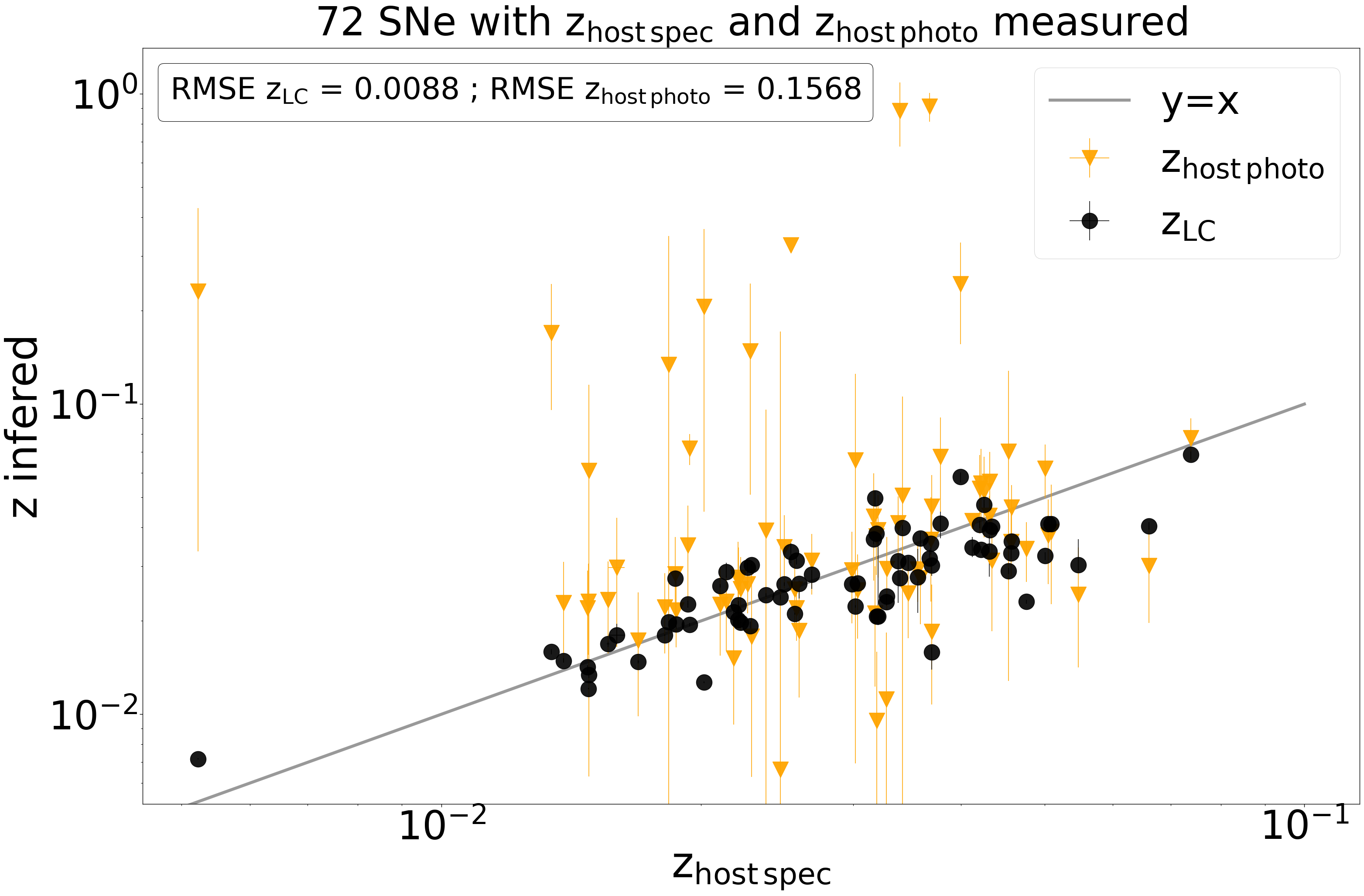}
	\caption{Relation between inferred redshift (z$_{\rm LC}$ and z$_{\rm host \, photo}$) with z$_{\rm host \, spec}$. Black circles correspond to z$_{\rm LC}$, while orange triangles correspond to z$_{\rm host \, photo}$. Error bars correspond to the errors (when reported) for host z, and for z$_{\rm LC}$ correspond to the lower and upper limits (percentiles 5 and 95 respectively). The root-mean square error for z$_{\rm LC}$ and z$_{\rm host \, phot}$ compared to z$_{\rm host \, spec}$ is reported at the top left corner.}
	\label{zinferedVSzspec}
\end{figure}

For the sample of 72 SNe whose host has z$_{\rm host \, spec}$ and z$_{\rm host \, photo}$ measurements available, we check how z$_{\rm LC}$ (black circles) and z$_{\rm host \, photo}$ (orange triangles) compare to z$_{\rm host \, spec}$ in Figure \ref{zinferedVSzspec}. We found that the RMSE when comparing z$_{\rm LC}$ to z$_{\rm host \, spec}$ is 0.0088, and the RMSE when comparing z$_{\rm host \, photo}$ to z$_{\rm host \, spec}$ is 0.1567. Also, we found that in 50 of the 72 cases, z$_{\rm LC}$ was closer to z$_{\rm host \, spec}$ value than z$_{\rm host \, photo}$. 

Our method could be used to estimate the redshift for type II SNe using only their LCs. Using type II SNe as distance indicators is something that previous works have tried with different methods \citep[see][]{KirshnerSNIIdistanc, HamuySNIIdistac, PoznanskiSNIIdistnace, Rodriguez2019SNIIdistanc, deJaegerSNIIdistanc}. The advantages of our method over the others are that it requires only a SN LC from any telescope; and it does not need the bolometric LC, spectroscopic information, or measuring features from the light curve.

\subsection{Sample distribution}

\begin{figure}[ht!]
	\centering
	\includegraphics[scale=.55]{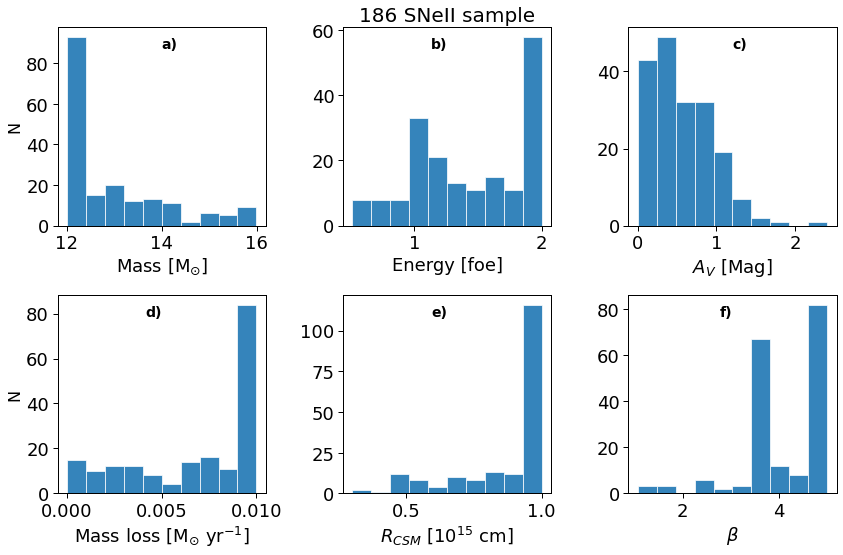}
	\caption{Parameter distribution histograms for the whole sample of 186 SNe II. a) Mass distribution. b) Energy distribution. c)  Attenuation A$_V$ distribution. d)  Mass loss rates distribution. e)Radii of the CSM distribution. f) $\beta$ from eq. \ref{beta_vel} distribution.}
	\label{params_dist}
\end{figure}

The median values shown in Table \ref{tabSNe_params}, are used to study the distribution of physical parameters for the whole sample of 186 SNe II. The marginalized distributions of physical parameters are shown in Figure \ref{params_dist}. Table \ref{tabsampledistribution} contains a summary of the main statistics of the marginalized distributions of parameters: the mean, standard deviation, median, percentile 5 (P$_5$), and percentile 95 (P$_{95}$).
 

  \begin{deluxetable*}{cccccc}


\tablecaption{Inferred parameters main statistics}
\tablenum{5}
\label{tabsampledistribution}

 \tablehead{
 \colhead{Parameter} & \colhead{Mean} & \colhead{$\sigma$} & \colhead{Median} & \colhead{P$_5$} & \colhead{P$_{95}$}}

 \startdata 
Mass [M$_\odot$]                       & 12.94  & 1.14   & 12.39  & 12.001  & 15.56    \\
Energy [foe]                           & 1.44   & 0.45   & 1.47   & 0.69    & 1.99     \\
Mass loss rate [M$_\odot$ year$^{-1}$] & 0.0068 & 0.0033 & 0.008  & 0.0007  & 0.0099   \\
r$_{CSM}$ [10$^{15}$ cm]                   & 0.87   & 0.17   & 0.98   & 0.50    & 0.99     \\
$\beta$                                & 4.18   & 0.83   & 4.15   & 2.50    & 4.99     \\
A$_V$ [mag]                            & 0.55   & 0.40   & 0.50   & 0.003   & 1.23     \\
 \enddata
 
 \tablecomments{$\sigma$ corresponds to one standard deviation. P$_5$ and P$_{95}$ correspond to the percentile 5 and 95 respectively.}
 
\end{deluxetable*}

Analyzing the sample distributions from Figure \ref{params_dist} we find that the mass distribution it is consistent with a power-law shape, which is consistent with the results in the literature for stars with mass $>$ 1 M$_\odot$ \citep{salpeter, IMF_Chabrier}. Our results also show that the models with a dense CSM that extends significantly above the star's photosphere are the ones that best represent our sample, i.e. the ones with a high CSM radius ($\sim 10^{15}$ cm), high mass loss rate ($>10^{-3}$ M$_\odot$/yr), and large $\beta$ value ($>$ 2). The double peak shape on the distributions of energy (near 1 and 2 foe) and $\beta$ (near 3.75 and 5), may be artifacts created by interpolating near the values of the grid of models.

\begin{figure}[ht!]
	\centering
	\includegraphics[scale=.45]{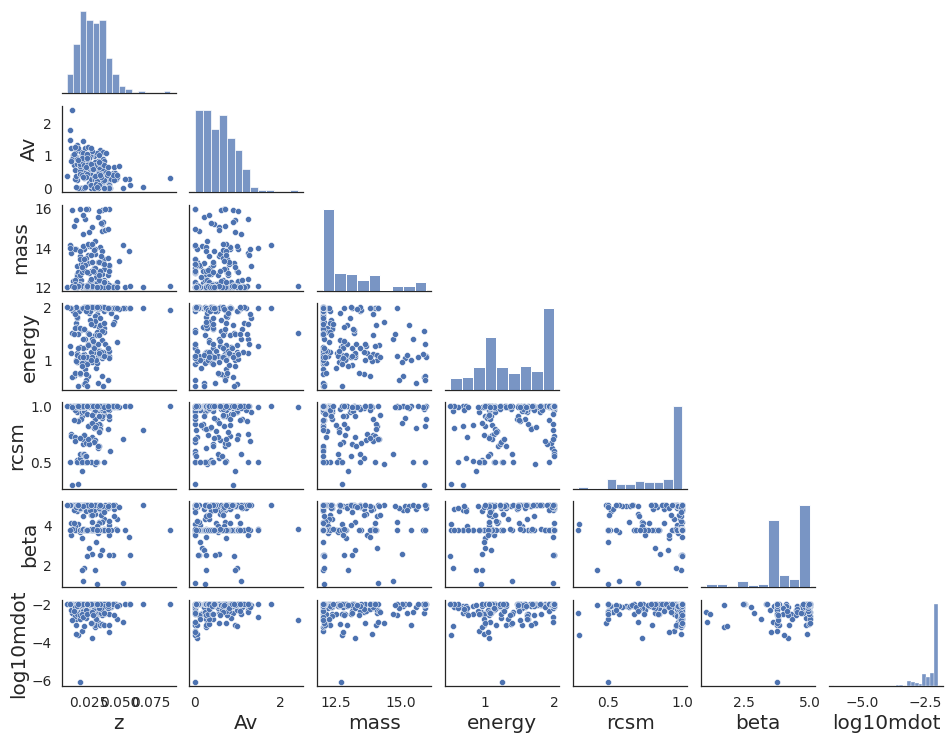}
	\caption{Pair-plot of the physical parameters distributions. From left to right (and from top to bottom): Redshift, Attenuation (A$_V$), mass [M$_\odot$], energy [foe], radius CSM [10$^{15}$ cm], $\beta$ velocity law value (from eq \ref{beta_vel}), and mass loss rate ($\dot{M}$) [M$_\odot$ yr$^{-1}$] in log scale.}
	\label{correlationplot}
\end{figure}

In Figure \ref{correlationplot} we present a pair-plot for the representative physical parameters. There are no clear visual correlations between any pair of parameters, aside from A$_V$ and redshift. However, it is worth noting that the apparent correlation between A$_V$ and redshift may be a result of degeneracy between these two parameters. Given the small z range for our sample, increasing A$_V$ and z have similar effects on the LC, making it fainter. This degeneracy could be resolved by adding filters at different wavelengths, as A$_V$ and z have different impact on the colors of the SNe. 

\begin{figure}[ht!]
	\centering
	\includegraphics[scale=.5]{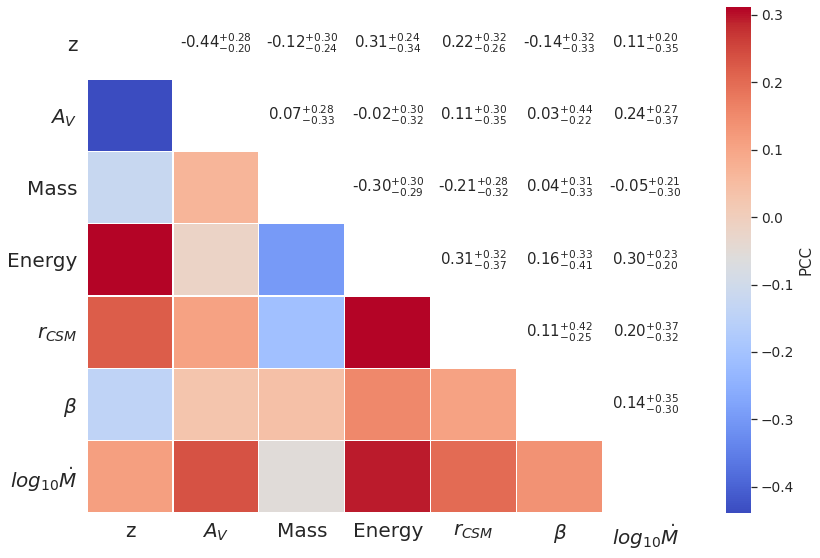}
	\caption{Correlation matrix between the physical parameters. The reported number corresponds to the Pearson correlation coefficient (PCC) and estimated error. From left to right (and from top to bottom): Redshift, attenuation, mass, energy, radius CSM, $\beta$ velocity law value (from eq \ref{beta_vel}), and mass loss rate.}
	\label{Correlation_matrix}
\end{figure}

We confirmed the apparent absence of correlations in Figure \ref{correlationplot} by computing the Pearson correlation coefficient (PCC) between parameters. The PCC measure the strength of the relationship of two variables and can take values between -1 and 1, where -1 means a perfect negative correlation, 1 a perfect positive correlation and 0 no correlation.
We also computed the PCC using bootstrapped samples with replacement 100 times for every combination of parameters. In Figure \ref{Correlation_matrix} we display a correlation matrix where the reported values correspond to the median of the PCC distribution and the subscript and superscript to the distances to the percentiles 2.5 and 97.5, respectively.

\subsection{Comparison with results from the literature} \label{sec:comparisonM22S22}

Now, we will discuss how our results compare to similar works, and its implications for future optical surveys, such as LSST. We would like to compare our results with independent measurements of any of the parameters inferred, as we did with the redshift. However, most of the other parameters are difficult to know without having direct information about the progenitor. Thus, our best option is to compare the distributions of physical parameters with those found in similar studies in the literature.

\begin{figure}[h]
	\centering
	\includegraphics[scale=.288]{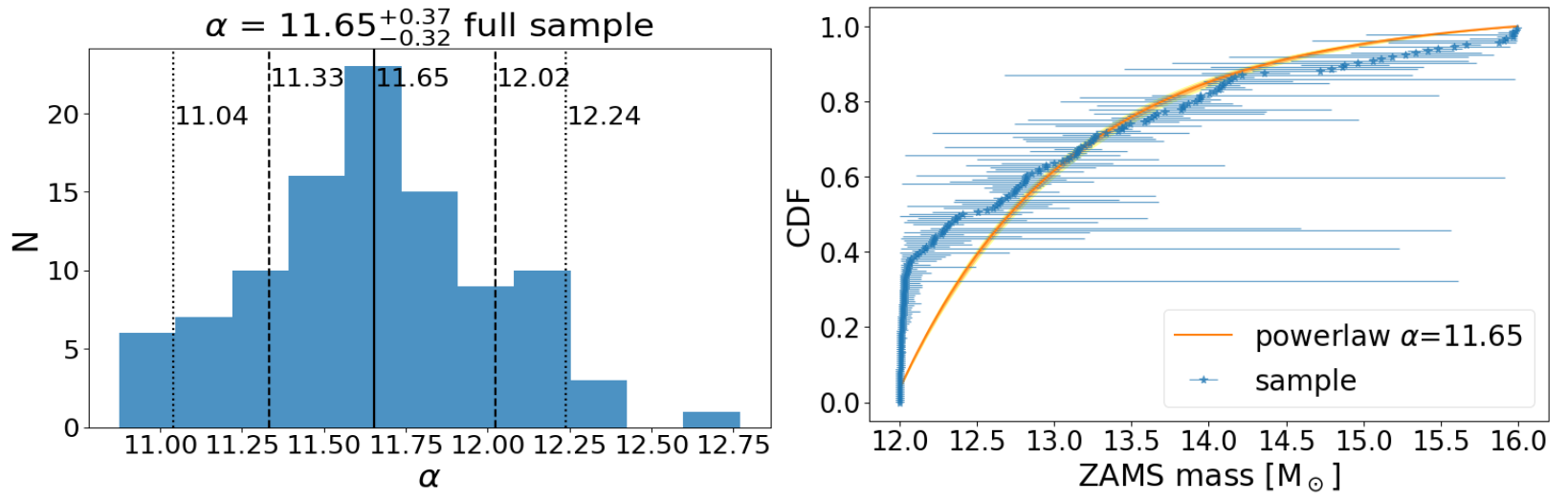}
	\caption{Mass IMF power-law exponent estimation for our sample. Left panel: The posterior distribution of $\alpha$. The median value is reported next to a continuous vertical line, the 16 and 84 percentiles are reported next to dashed vertical lines, and the 5 and 95 percentiles are reported next to dotted vertical lines. Right panel: Cumulative distribution function (CDF) of our mass distribution and CDF of a power law distribution with an exponent equal to the median value reported on the left panel. The orange surface and yellow surface represents the 1 $\sigma$  and 2 $\sigma$ error respectively}
	\label{AlphaWholeSample}
\end{figure}

\begin{figure}[h]
	\centering
	\includegraphics[scale=.29]{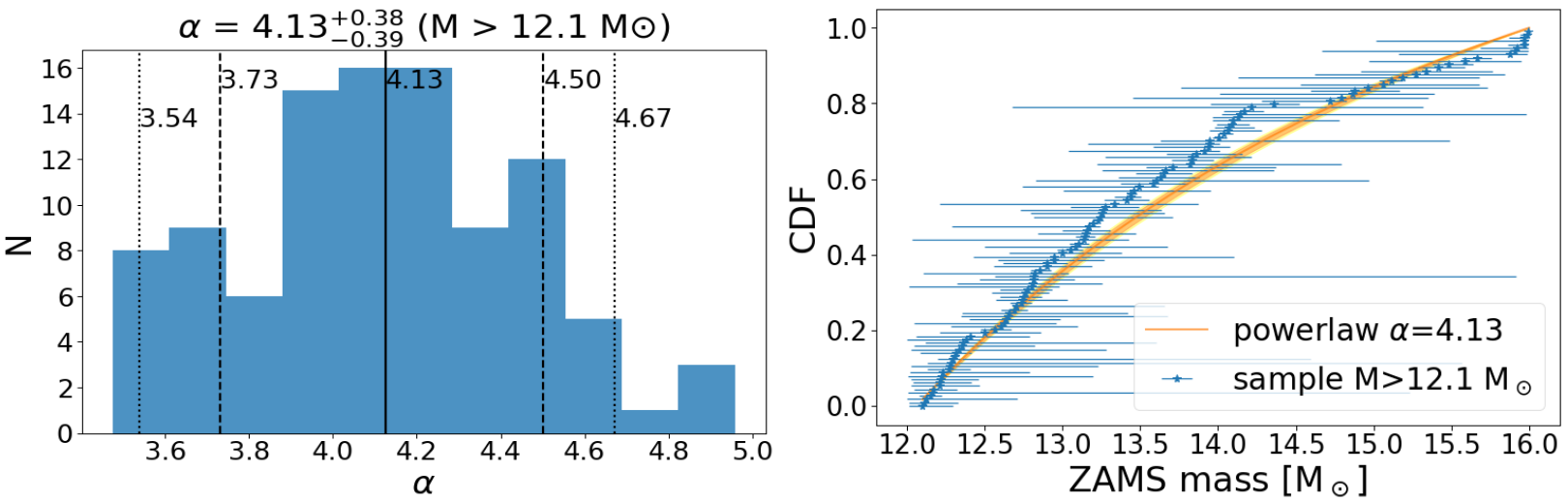}
	\caption{Same as Figure \ref{AlphaWholeSample} but for SNe with inferred masses $\geq 12.1 M_\odot$ }
	\label{Alpha12.1}
\end{figure}

\citetalias{Martinezcorr} inferred SNe II physical parameters from bolometric LCs and photospheric velocities using hydrodynamical models and studied the correlation between physical and observed parameters from a sample of SNe II from the Carnegie Supernova Project-I. In their work, they found a weak correlation between explosion energy and ZAMS mass of the progenitor star that differs from the weak (or negligible) negative correlation we found for these two parameters. They found that the inferred masses from their sample followed a power-law with exponent $\alpha$ = 4.07$^{+0.29}_{-0.29}$ for their whole sample, and $\alpha$ = 6.35$^{+0.57}_{-0.52}$ for their gold sample. Both values are steeper than the Salpeter IMF. This was called the IMF incompatibility by \citetalias{Martinezcorr}. They concluded that this incompatibility is due to the lack of understanding of some physical ingredients and not related to the completeness of their sample.
Using a Bayesian approach we found a value of $\alpha$ = 11.65$^{+0.37}_{-0.32}$ (see Figure \ref{AlphaWholeSample}) steeper than the values found in \citetalias{Martinezcorr}. Our inferred value of $\alpha$ could be overestimated as a consequence of our limited parameter space. Our mass parameter space goes from 12 M$_\odot$ to 16 M$_\odot$ in contrast with \citetalias{Martinezcorr} which goes from 9 M$_\odot$ to 25 M$_\odot$. In Figure \ref{AlphaWholeSample} it is possible to see how our limited parameter space affects our estimation of $\alpha$ as most of our masses are stacked near $12 M_\odot$. To avoid this issue we repeat the analysis but with the SNe in our sample with inferred masses greater than $12.1 M_\odot$. We obtained a value of $\alpha$ = 4.13$^{+0.38}_{-0.39}$ (see Figure \ref{Alpha12.1}) comparable with the result of \citetalias{Martinezcorr} for their whole sample.

It is worth noting that \citetalias{Martinezcorr} used the measured photospheric velocity to avoid degeneracy in the parameter estimation, i.e that two different models produce the same bolometric LC. Another significant difference between our work and \citetalias{Martinezcorr} lies in the range of energies explored. Our work energy grid goes from 0.5 to 2 foe, while \citetalias{Martinezcorr} grid goes from 0.1 to 1.5 foe. This discrepancy could potentially impact the inferred values, given that \citetalias{Martinezcorr} encounters energy values peaking around 0.5 foe, while our distribution peaks at the upper limit of our models, 2 foe. This discrepancy might be attributed to a combination of factors. Firstly, not including photospheric velocities in our analysis. Secondly, the data used on \citetalias{Martinezcorr} does not take into account the early part of the LC as constraint as in our work. Therefore, this discrepancy appears to have arisen because high energy is required to constrain the rapid rise of the early part of the SNII LC, while lower energies are needed to explain the measured photospheric velocities. Further analysis beyond the scope of our work is needed to fully understand this difference.

\begin{figure}[h]
	\centering
	\includegraphics[scale=.39]{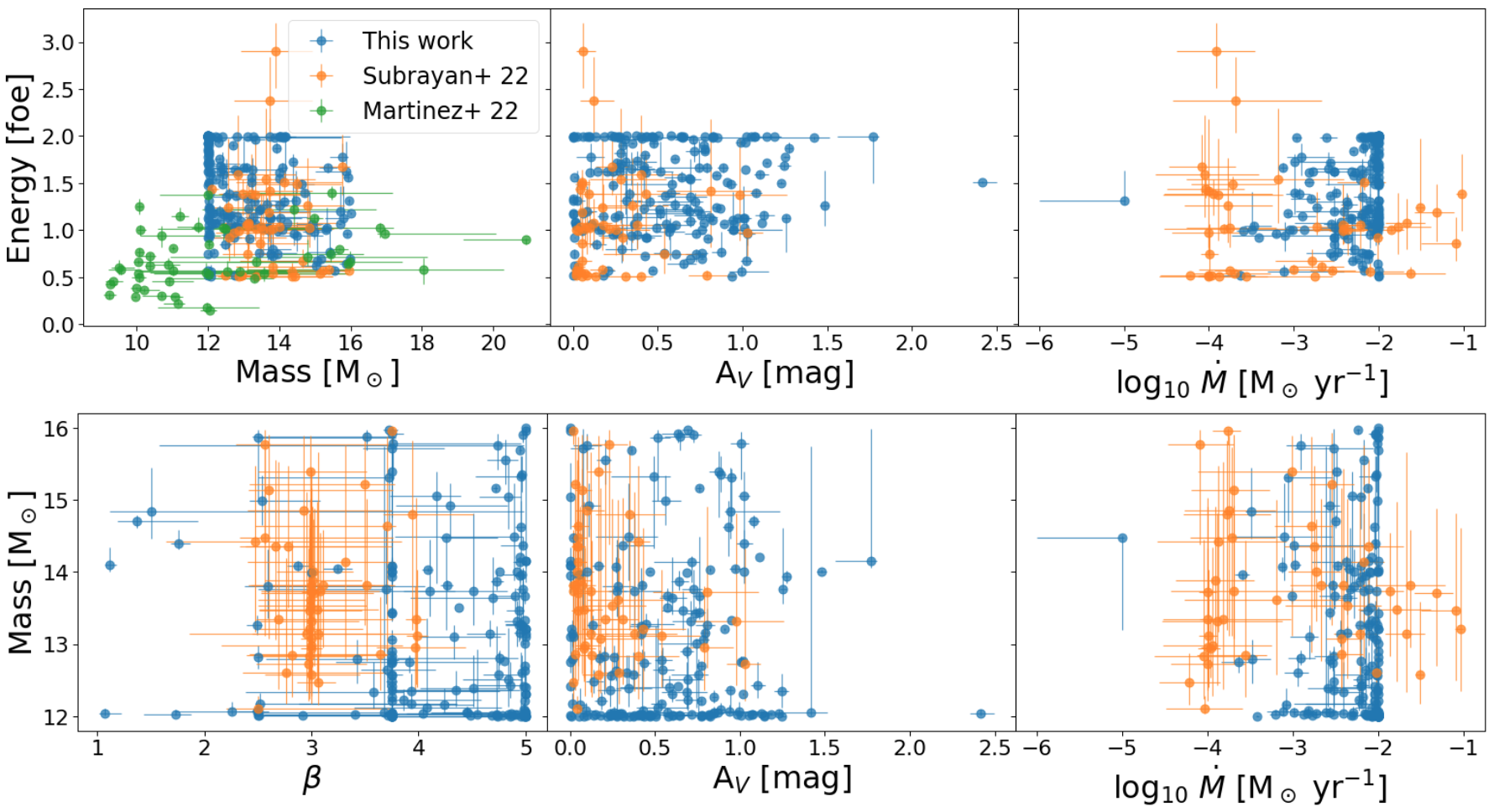}
	\caption{Combination of parameters comparison of our results (blue circles), \citetalias{Subrayan} results (orange circles), and \citetalias{Martinezcorr} results (green circles). Top panel: Energy plotted against ZAMS mass, attenuation (A$_V$), and mass loss rate. Bottom panel: ZAMS mass plotted against $\beta$ parameter from the velocity law,  attenuation (A$_V$), and mass loss rate.}
	\label{scattercomparisonS+M}
\end{figure}

\citet[][hereafter \citetalias{Subrayan}]{Subrayan} used wind-enhanced models from \cite{TakashiNewModels}, similar to our work, to infer physical parameters from a sample of 45 SNe. The main differences with our analysis is that they use only ZTF data without including other surveys, they include $^{56}$Ni mass as a parameter of the model, they did not infer the redshift nor the CSM radius, and they explored a significantly larger parameter space for energy and mass loss rates. They focused on the effect of using hydrodynamical modeling to forecast and guide follow-up observations as the light curve of the SNe evolves in the context of the Recommender Engine For Intelligent Transient Tracking \citep[REFITT][]{2020ApJ...893..127S}. In their work they find $\beta$ values around 3 and mass loss rates between 10$^{-4}$ M$_\odot$ yr$^{-1}$ - 10$^{-2}$ M$_\odot$ yr$^{-1}$ consistent with our results. They also could not find a significant correlation among the inferred parameters. A comparison between their results, \citetalias{Martinezcorr}, and our results is shown in Figure \ref{scattercomparisonS+M} where we show the distributions of energy with mass, A$_V$, and mass loss rate; and mass with $\beta$, A$_V$, and mass loss rate. It is possible to see that our mass parameter space is smaller than \citetalias{Martinezcorr} and \citetalias{Subrayan}. Overall we found similar trends regarding the preference for lower mass stars and dense CSM models between our work \citetalias{Martinezcorr}, and \citetalias{Subrayan}. However, in \citetalias{Subrayan} they infer mass loss rates that are inside our grid of values and significantly below our inferred values for a large fraction of their sample. 

\begin{figure}[h]
	\centering
	\includegraphics[scale=.30]{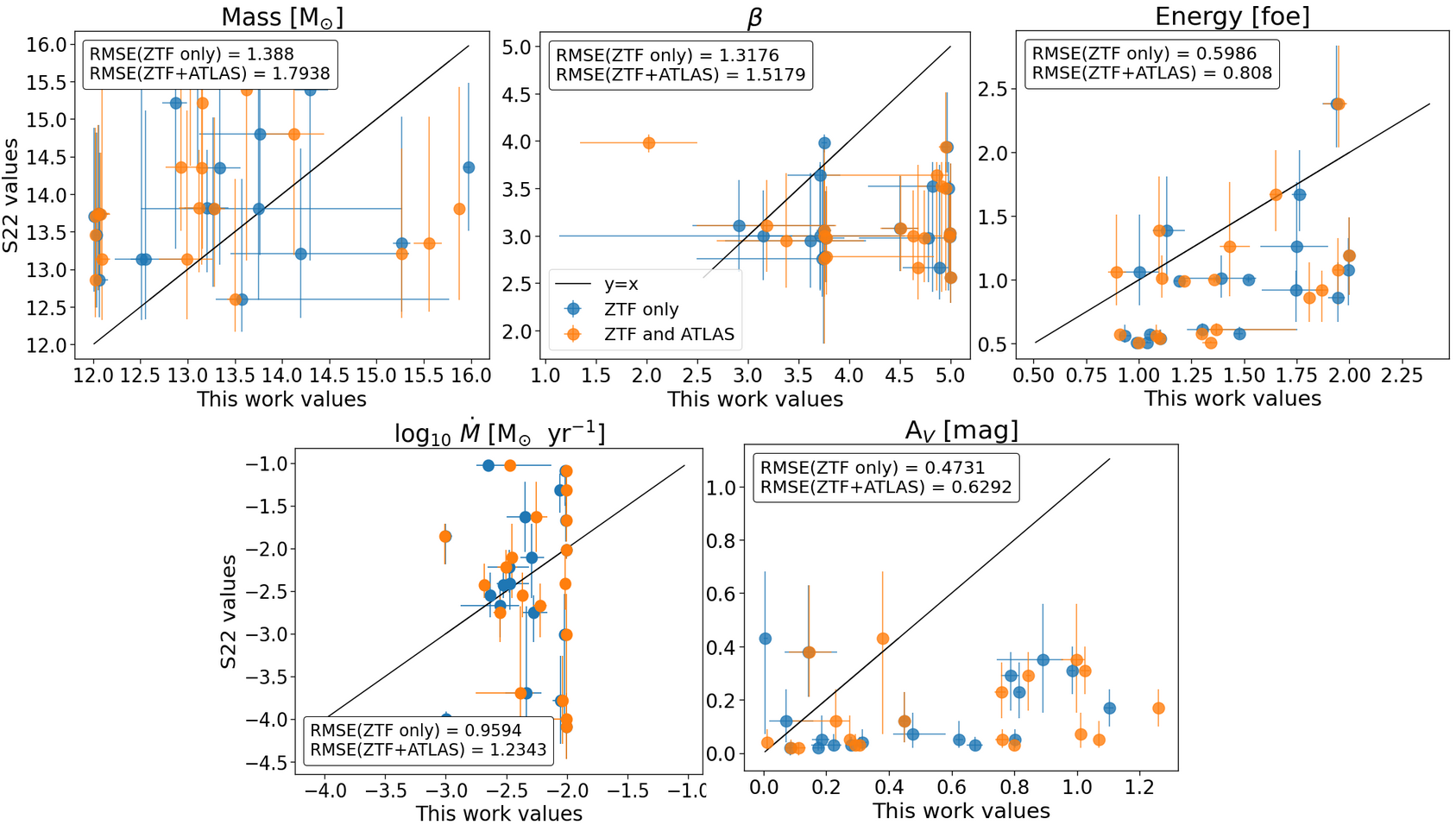}
	\caption{Comparison of the inferred parameters for 19 SNe in both our and \citetalias{Subrayan} samples with our method using ZTF and ATLAS forced photometry with redshift fixed (orange dots), and with our method using only ZTF forced photometry with redshift fixed (blue dots). The top panel, from left to right, corresponds to mass, wind acceleration parameter $\beta$, and energy. The bottom panel corresponds to the mass loss rate and attenuation. RMSE calculated as the mean square root of the difference between the values of \citetalias{Subrayan} and our work is reported for every parameter for our method using ZTF and ATLAS data, and ZTF data only.}
	\label{s22comparison}
\end{figure}

A subset of 19 SNe were studied in both our work and \citetalias{Subrayan}. A comparison between our inferred parameters and those from \citetalias{Subrayan} is shown in Figure \ref{s22comparison}. It is important to note that the comparison in the previous analysis could be influenced by the fact that \citetalias{Subrayan} did not include ATLAS forced photometry in their analysis. To address this potential influence, we estimate the parameters of the 19 SNe from the two samples using both our original results and trying to mimic their results doing the inference again using only the ZTF information. Despite finding similar overall trends in the distributions of parameters as previously mentioned, Figure \ref{s22comparison} shows that the same sample of SNe leads to different inferred values between our work and \citetalias{Subrayan}.

We believe that the disagreement may be related to the differences in the grid of models and posterior sampling method used. It is not entirely clear how the explosion times are determined in \citetalias{Subrayan}, which may also explain some of the differences found. A combination of higher energy and lower mass produces fast-rising LCs. \citetalias{Subrayan} has a larger energy space to explore. Therefore, some of their SNe can have similar rise times to ours having larger masses with higher energies. We previously mentioned that the double peak shape found in our energy and $\beta$ distributions could be an artifact due to the values of our models. It is noteworthy that \citetalias{Subrayan} could encounter a similar situation, with values of $\beta$ close to 3 (see Figure \ref{scattercomparisonS+M}), which may affect the inference for the other parameters and thus cause the difference seen in Figure \ref{s22comparison}. Although the reported RMSE after trying to mimic their inference diminishes for all the parameters, there are still large discrepancies between our work values and \citetalias{Subrayan}. A more detailed comparison is required to understand these differences.

\subsection{Validation using simulated LCs}
In this subsection, we aim to address the discrepancies observed in the previous subsection, particularly in the case of \citetalias{Subrayan}. Our goal is to test the reliability of our method and confirm whether any differences are due to differences in the methods or limitations in our approach.

We selected a subsample of 20 random SNe from our dataset. We simulated LCs for these SNe by interpolating the models using the parameters we had inferred and considering the cadence specific to each SN. Subsequently, we run our method on these simulated LCs. We tried three different configurations to run our method. First, we started with initial values around the middle of the parameters possible range and used 900 steps. Second, we did the same but with 2000 steps. The third configuration was the one we used in our study and \citetalias{forster2018delay}, where we started with an educated guess or the parameters and also used 900 steps by running an interactively fitting routine.

\begin{figure}[h]
	\centering
	\includegraphics[scale=.272]{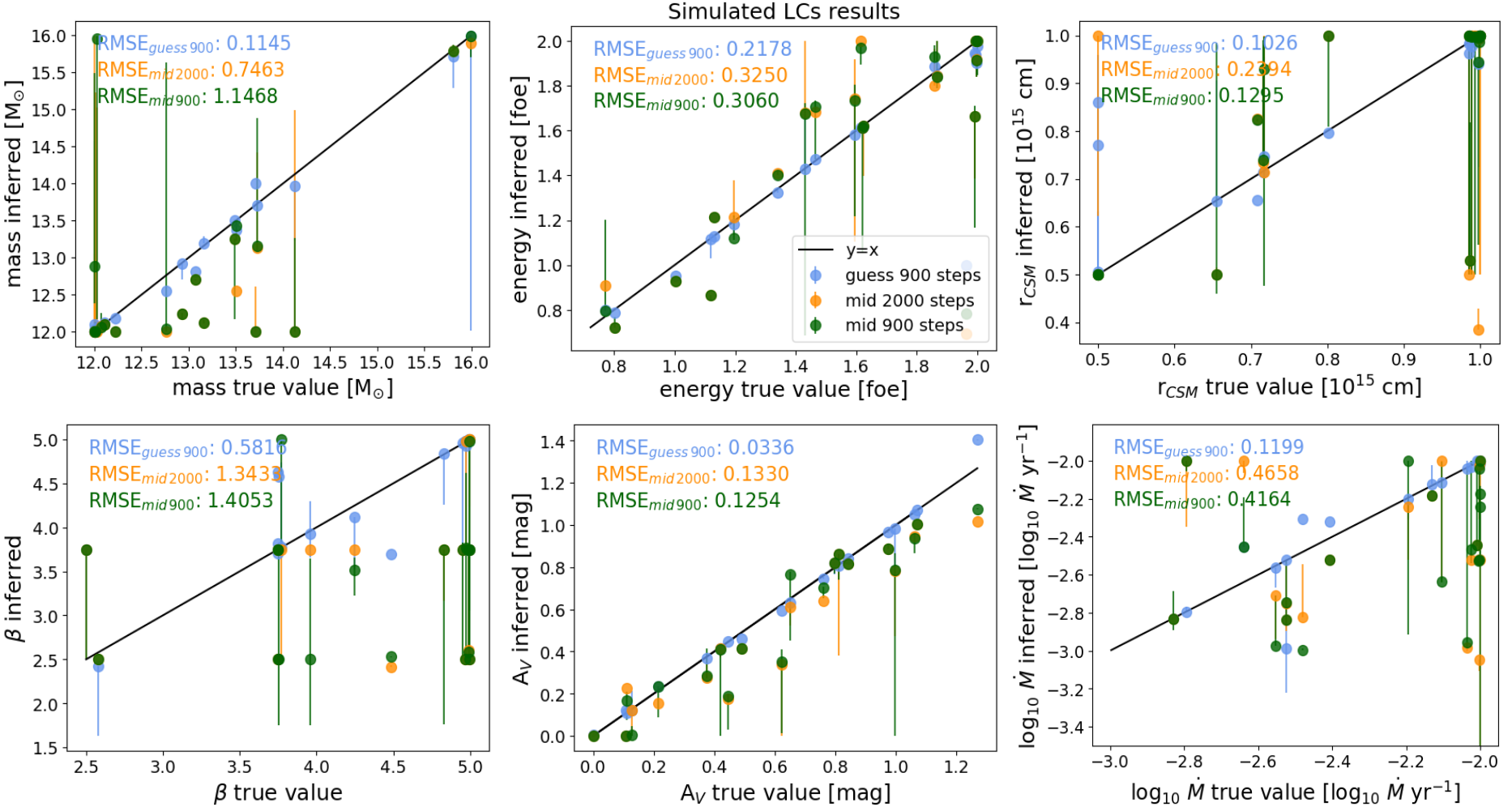}
	\caption{Comparison of inferred values from simulated LCs with true simulated values. Blue dots corresponds to walkers initialized near the MAP with 900 steps, orange dots corresponds to walkers initialized around the middle of the parameter space for every variable with 2000 steps, green dots indicates the same as orange dots, but with 900 steps. The RMSE for each parameter and configuration are reported at the top left for every parameter. Top Row (left to right): mass, energy, r$_{CSM}$. Bottom Row (left to right): $\beta$, attenuation, mass loss rate.}
	\label{Simulation_results}
\end{figure}

The results of the inferred values for every configuration are shown in Figure \ref{Simulation_results}, displaying inferred values versus true values for the three different configurations. Points close to the identity line (y=x) indicate close agreement between inferred and true values. We observe that the method with initialization close to the Maximum A Posteriori (MAP) estimate produces significantly superior results compared to starting from the middle of the grid, even with increased steps, for all parameters. Specifically, we find that Energy and A$_V$ are consistently mostly well reproduced across all configurations. On the other hand, mass and mass loss are accurately reproduced only for the initialized configuration; without initialization, they are mostly underestimated. Lastly, accurately obtaining values for r$_{\rm CSM}$ and $\beta$ remains challenging across all configurations. In Table \ref{tabsimulations}, we summarize the RMSE of the different inferences of parameters for the three different configurations and for the comparison with \citetalias{Subrayan}. We do not observe a significant improvement in parameter inference when employing a larger number of MCMC steps. More importantly, we find that the initial estimation plays a critical role in obtaining accurate results. The importance of doing a good initialization has been discussed in previous works such as \citet{2013PASP..125..306F} and \citet{2018ApJS..236...11H}, where the advantages of initializing the walkers around the expected MAP are mentioned in order to prevent walkers from converging to lower probability modes of the posterior distribution and to accelerate convergence.

\begin{deluxetable*}{cccccccc}


\tablecaption{Simulations RMSE}
\tablenum{6}
\label{tabsimulations}

 \tablehead{
 \colhead{Parameter} & \colhead{MAP 900 steps} & \colhead{Mid 900 steps} & \colhead{Mid 2000 steps} & \colhead{\citetalias{Subrayan} vs ZTF+ATLAS} & \colhead{\citetalias{Subrayan} vs ZTF only}}

 \startdata 
 mass [M$_\odot$] & 0.114 & 1.146 & 0.746 & 1.793 & 1.388 \\
 energy [foe] & 0.217 & 0.306 & 0.325 & 0.808 & 0.598 \\ 
 log$_{10}$$\dot{M}$ [M$_\odot$ yr$^{-1}$] & 0.119 & 0.416 & 0.465 & 1.234 & 0.959 \\
 r$_{CSM}$ [10$^{15}$ cm] & 0.102 & 0.129 & 0.239 &  &  \\ 
 A$_V$ [mag] & 0.033 & 0.125 & 0.133 & 0.629 & 0.473 \\ 
 $\beta$ & 0.581 & 1.405 & 1.343 & 1.517 & 1.317
 \enddata

\tablecomments{MAP refers to the initialization using an interactive fitting close to the Maximum a posteriori (MAP), while Mid refers to the initialization close to the middle of the parameter space for every variable. \citetalias{Subrayan} vs ZTF+ATLAS and \citetalias{Subrayan}  vs ZTF only correspond to the RMSE between the \citetalias{Subrayan} reported values and our inferred values ZTF and ATLAS forced photometry or only ZTF forced photometry, respectively.}


\end{deluxetable*}

The interactive fitting to establish an initialization for MCMC significantly enhances the accuracy of inferred parameters as shown in Table \ref{tabsimulations}. This suggests that the discrepancies between our results and those of \citetalias{Subrayan}, as discussed in the previous section, seems to be related to differences in the methods rather than inaccuracies in our approach, this is supported by the fact that the RMSE from our simulations is lower than that observed in the comparison with \citetalias{Subrayan} . Furthermore, these discrepancies could be significantly influenced by variations in the initialization methods and how each method handles multimodality.

\subsection{Limitations of our method}

Currently, our method is constrained by several limitations. Firstly, the models employed in our study explore only a confined region within the parameter space (see Table \ref{tabparams}). For example, we lack SN models originating from progenitor stars with masses below 12 M$_\odot$. Additionally, there are physical parameters effects not considered in this analysis, such as the variation of $^{56}$Ni mass, capable of influencing SN light curves. 

A limitation for future LSST applications of the proposed method is that it needs SNe that are well observed during the rise and at the peak of the LC. If we do not have information that can constrain the explosion time of the SN, our results can be extremely inaccurate. This is why we discarded 70 SNe from our sample of 256 confirmed type II SNe. Also, any gap in the LC could mean a bimodality in some parameters or inaccurate results as shown in Figure \ref{posterior_fewpoints}, where the lack of data points between MJD 58920 - 59050 in the LC of SN 2020aer produces wider distributions, i.e. less precise results, in the posterior. Furthermore, our method uses significant computational resources. We use the \verb+largemem+ partition from the National Laboratory for High Performance Computing (NLHPC\footnote{\url{https://www.nlhpc.cl/}}) where we need to allocate 5 Giga Bytes of memory to be able to run the code for one SN. This is due to our preloading of synthetic light curves during code execution. Therefore, before expanding the number of models, optimizing memory utilization becomes fundamental. Failure to do so could lead to memory requirements exceeding our available resources.

\subsection{Implementation on LSST }

As stated previously, the number of SNe that will be discovered in the LSST era is going to be an order of magnitude larger than what we are experiencing today, so it is necessary to be prepared for this challenge. Our current method inferred physical parameters from a sample of 186 SNe within less than 12 hours using the NLHPC. Looking forward to the LSST era, our method could be in principle adapted to use with LSST data, but significant work is needed to fix the limitations discussed in the previous section. A possible solution is to use a surrogate model to make the likelihood computation faster and more memory efficient, e.g. using neural networks.

One significant challenge we anticipate when applying our method to LSST data is the cadence, that initially will not be as fast as the ZTF cadence (2-3 night cadence). Based on our experience gained in this study, the ZTF cadence proved sufficient in some cases for accurately inferring the physical parameters solely from ZTF data. However, in other cases, the ZTF data alone was insufficient and ATLAS data, that has a 2 night cadence, was necessary. Otherwise, we could have multi modality or incorrect posterior distributions. Therefore, we believe that it will be necessary to complement LSST data with data from other facilities such as another survey or follow-up observations. The latter seems more difficult given the volume of SNe to be discovered, so in this work we use data from two surveys, ZTF and ATLAS. One of the advantages of LSST, besides the amount of SNe it will discover, is going to be the six available filters, that will allow our method to constrain better the value of A$_V$ and thus a better redshift estimation and overall normalization of the light curve. An additional advantage of LSST, is its ability to precise constraints on the $^{56}$Ni mass from the tail luminosity. The capability of LSST to constrain this parameter for a larger number of SNe is notable, and in some cases, with good cadence, such as SNe in the deep drilling fields, it will offer very good constraints on the explosion epoch


Finally, the fact that our method can infer the redshift better than z$_{\rm host \, photo}$ is promising. Given that a considerable amount of SNe will have a host galaxy whose redshift has not been measured, it will also allow photometric redshift campaigns to have an independent redshift to compare with.

\section{Conclusion} \label{sec:Conclusion}

We have developed a method that can be used in data from any photometric survey to infer physical parameters of SNe type II using the models from M18, which is $\sim$ 6 times faster than \citetalias{forster2018delay}. The method was applied to a sample of 186 SNe in less than 12 hours using the NLHPC computer cluster. We studied the distribution of physical parameters and found that the dense CSM models are the ones that best represent SNe type II. We found a low negative correlation between redshift and attenuation; however, it is important to note that this correlation is not physically meaningful but rather a degeneracy between these two parameters. Other parameters show a negligible correlation as seen in the correlation matrix in Figure \ref{Correlation_matrix}. In the process, we developed a method and guidelines to clean forced photometry data from the ZTF and ATLAS forced photometry services for transient object LCs.


We compare our light curve inferred redshifts (z$_{\rm LC}$) with the host galaxy redshifts for those SNe where this was available and found that our method is capable of estimating the redshift for a SN with better accuracy than the host photometric redshift. Considering the amount of data that LSST will produce, we can use our method to estimate the redshift of SNe II based on the LCs. However, to implement this we assume that the sample of SNe II will be well classified and with no significant contamination from other classes, which highlights the importance of photometric classification provided by LSST community Brokers.

We use Bayesian inference to estimate the exponent from a power-law distribution that fits the distribution of inferred progenitor masses. We find a value of $\alpha$ = 11.65$^{+0.37}_{-0.32}$, steeper than the Salpeter IMF value ($\alpha$ = 2.35). We did the same analysis but only for SNe in our sample with an inferred mass $>$ 12.1 M$_\odot$ and find a value of $\alpha$ = 4.13$^{+0.38}_{-0.37}$ similar to the one found in \citetalias{Martinezcorr} for their gold sample. Despite having a result in agreement with \citetalias{Martinezcorr}, a larger grid of models in mass is needed to more confidently derive $\alpha$.

We conduct a comprehensive comparison of our method with two studies from the literature, \citetalias{Martinezcorr} and \citetalias{Subrayan}. While our mass distribution results exhibit similarities with those of \citetalias{Martinezcorr}, we find differences in the energy distribution. We observe higher energies than \citetalias{Martinezcorr}. This may be a consequence of not including photospheric velocities or using a more restrictive light curve for the early part. Further analysis is necessary to understand this difference.
A noticeable discrepancy emerged when comparing our results with \citetalias{Subrayan}. Despite employing similar families of models, the discrepancies may arise due to differences in the grid of models, the input data that was used, and/or due to variations between the Bayesian inference methodologies employed by \citetalias{Subrayan} and our approach. We infer parameters using ZTF data, as \citetalias{Subrayan} did, to examine whether the discrepancies were a result of differences in the input data. Despite using only ZTF data, we still observe discrepancies. This finding indicates another factor is responsible for the discrepancies. 

We performed simulations to test the robustness of our method. We find that the initialization is crucial for our method to infer parameters correctly. The discrepancies with \citetalias{Subrayan} seem to arise from methodological differences, such as initialization; Bayesian approach; and grid of models, rather than inherent inaccuracies, although a more detailed comparison may be needed Therefore, caution should be used when working with hydrodynamical models to infer physical parameters. We need to better understand the influence of the grid of models, the quality of the input data, the regions in the light curves that better constrain some of the parameters, and whether the inference method is well calibrated. This understanding may help us distinguish true variations due to physical differences that may lead to new insights.

Most SNe discovered by LSST will not have spectroscopic classification. Therefore to use our method with SNe with no spectral information, photometric classification will be necessary \citep[e.g.][]{forster2018, alerce_general, ALeRCE_LC}. Also, given the cadence of LSST, data from other telescopes may be necessary to complement the LCs and reduce the uncertainty in the posteriors.

Our method uses a large amount of memory when running. Therefore, more optimization is needed. A possible solution for this problem could be to implement our method in a faster programming language or using other approaches to infer the posterior such as Amortized Variational Inference \citep{sanchez2021amortized, 2022arXiv221104480V}. Lastly, looking for another way of representing the synthetic LCs could save memory usage.

Our method is flexible, so it can be used in other models, as long as synthetic time series of spectra are available. We look forward to testing it with different models, or with more enhanced wind scenario models so we can explore a bigger parameter space. Also, more models could produce a more refined model grid that will allow for more accurate interpolation.

The code used in this work is publicly available in \url{https://github.com/fforster/surveysim/tree/dev-javier}.

\section{Acknowledgments}

 Powered@NLHPC: this research was partially supported by the supercomputing infrastructure of the NLHPC (ECM-02). We acknowledge support from FONDECYT REGULAR 1200710 (JSF, FF). This work was funded by National Agency for Research and Development (ANID), Millennium Science Initiative ICN12$\_$009 (FF, LHG, AMMA, JPA, AC). FF and AMMA acknowledge the BASAL Center of Mathematical Modelling Grant PAI AFB-170001. This work uses data from the University of Hawaii's ATLAS project, funded through NASA grants NN12AR55G, 80NSSC18K0284, and 80NSSC18K1575, with contributions from the Queen's University Belfast, STScI, the South African Astronomical Observatory, and the Millennium Institute of Astrophysics, Chile. The ZTF forced-photometry service was funded under the Heising-Simons Foundation grant $\#$12540303 (PI: Graham). The NASA/IPAC Extragalactic Database (NED) is funded by the National Aeronautics and Space Administration and operated by the California Institute of Technology. We thank the ALeRCE Broker for making their services public to the scientific community. In this work we used their Web Interface and the ZTF Forced Photometry Notebook.

\appendix

\section{Model's grid interpolation} \label{model_interpolation}

To be able to interpolate quickly between models with different physical parameters \citetalias{forster2018delay} introduced the following interpolation.

For a set of parameters $\vec{\theta}$ we start by finding the closest values in all the intrinsic physical dimensions and find all the models that have combinations of these values, which is called $\vec{\theta}_{close}$. The final LC will be a weighted combination of all these models

\begin{equation}
m(t, t_{exp}, z, A_{V} ,\vec{\theta}) = \sum_{\vec{\theta}_{i} \in \vec{\theta}_{close}} \hat{w}(\vec{\theta}, \vec{\theta}_{i}) m((t, t_{exp}, z, A_{V} ,\vec{\theta}_{i}),
\end{equation}

where $m(t, t_{exp}, z, A_{V}, \vec{\theta})$ is the magnitude of the model at a given observation time $t$, explosion time $t_{exp}$, redshift $z$, a given attenuation $A_V$ and a given vector of parameters $\vec{\theta}$. $w$ are the normalized weights that are defined as:

\begin{equation}
\hat{w}(\vec{\theta}, \vec{\theta}_{i}) = \frac{w(\vec{\theta}, \vec{\theta}_{i})}{\sum_{\vec{\theta}_{j} \in {\vec{\theta}}_{close}} w(\vec{\theta},\vec{\theta}_{j})},
\end{equation}
where the weights $w$ are defined to be inversely proportional to the product of the differences of the vector of physical parameters $\vec{\theta}$

\begin{equation}
w(\vec{\theta},\vec{\theta}_{i}) = (\prod_{j} |\vec{\theta}^{j} -\vec{\theta}^{j}_{i}| + \delta^{j})^{-1}
\end{equation}
where $\vec{\delta}$ has the same units as $\vec{\theta}$, but much smaller than the separation of the grid, in order to ensure that the weights do not diverge.

\section{Host galaxy association} \label{Appendix_hostassoc}

\begin{figure}[ht!]
	\centering
	\includegraphics[scale=.41]{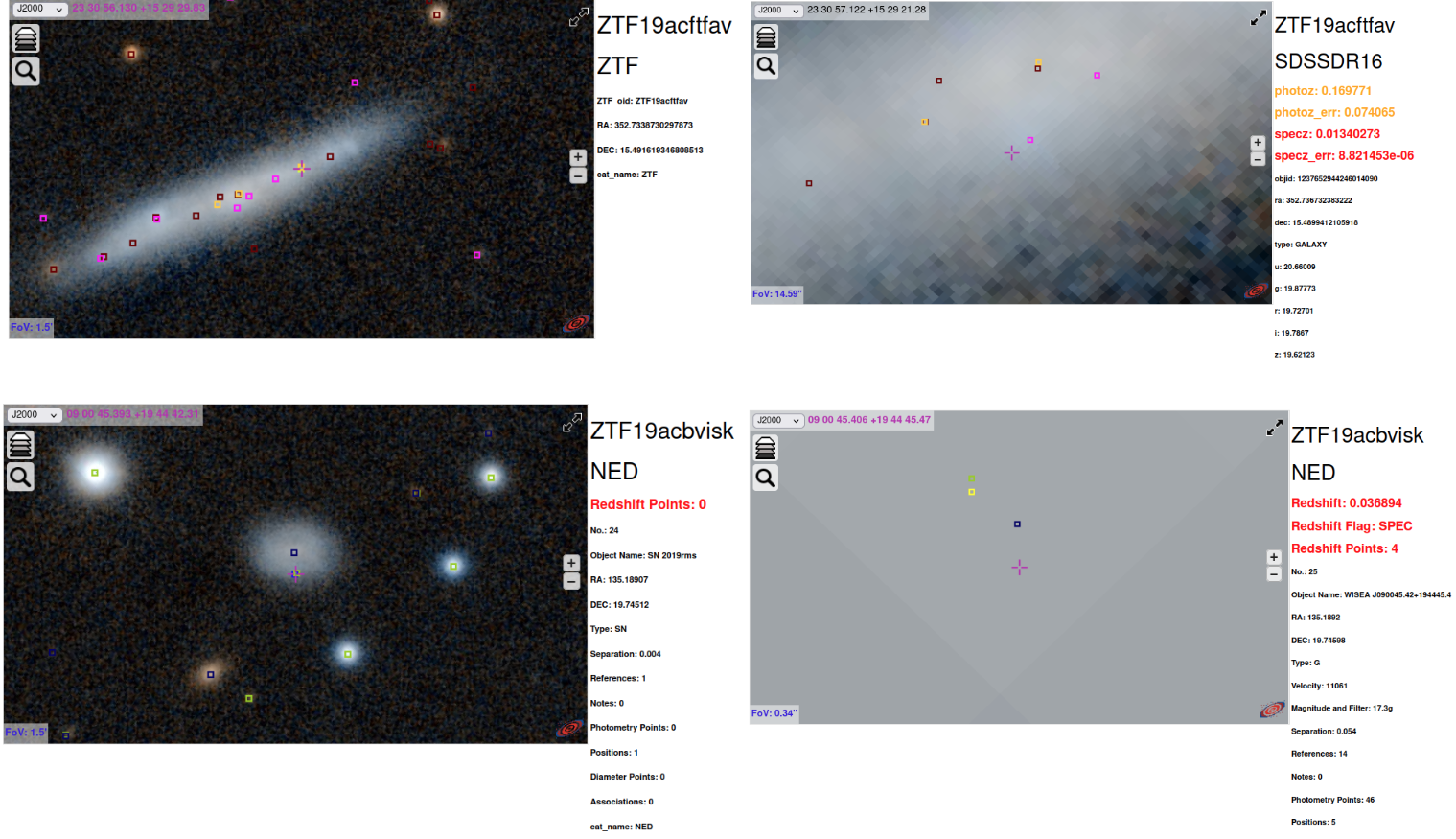}
	\cprotect\caption{Left panels: Interactive \verb|ipyladin| panels of Pan-STARRS HiPS images with a field of view (FOV) of 1.5' centered in the SN location. The multiple squares correspond to the locations of known sources in NED, SIMBAD and SDSS DR16. The top row corresponds to ZTF19acftfav and the bottom row corresponds to ZTF19acbvisk. Right panels: Same as left panels but zoomed in on the core of the host galaxy. Information about each known source is included next to the images, and is determined by the position of the cursor (see text for more details).}
	\label{host_assoc_plot}
\end{figure}

To determine the host galaxy we do a visual inspection using \texttt{\textbf{ipyaladin}} ipyaladin \citep{ipyaladin}. The process of associating each SNe to a host is as follows: 1) we load a Pan-STARRS-DR1 image from bands z and g as seen in the left panels of Figure \ref{host_assoc_plot}, where every color square is a source in the NED, SIMBAD, or SDSS DR16 catalogs. We can hover the cursor over the different squares to look for information about the source, such as the name of the source, the survey, the redshift of the object (if available), the kind of redshift (spectroscopic or photometric), etc. We visually associate a SN to its host galaxy, and if multiple sources (squares in Figure \ref{host_assoc_plot}) are in the core of the host, we select the one with the best redshift available (spectroscopic over photometric), and if possible the best available that report errors. 2) We select the source by clicking on the corresponding square and saving the following information about the host: name, right ascension, declination, offset from SN coordinates, source catalog name, redshift spec flag (True if redshift is spectroscopic, False otherwise), redshift, redshift error, and redshift type.

\section{Supplementary Figures} \label{Supplement_Figures}

\begin{figure}[h!]
	\centering
	\includegraphics[scale=.32]{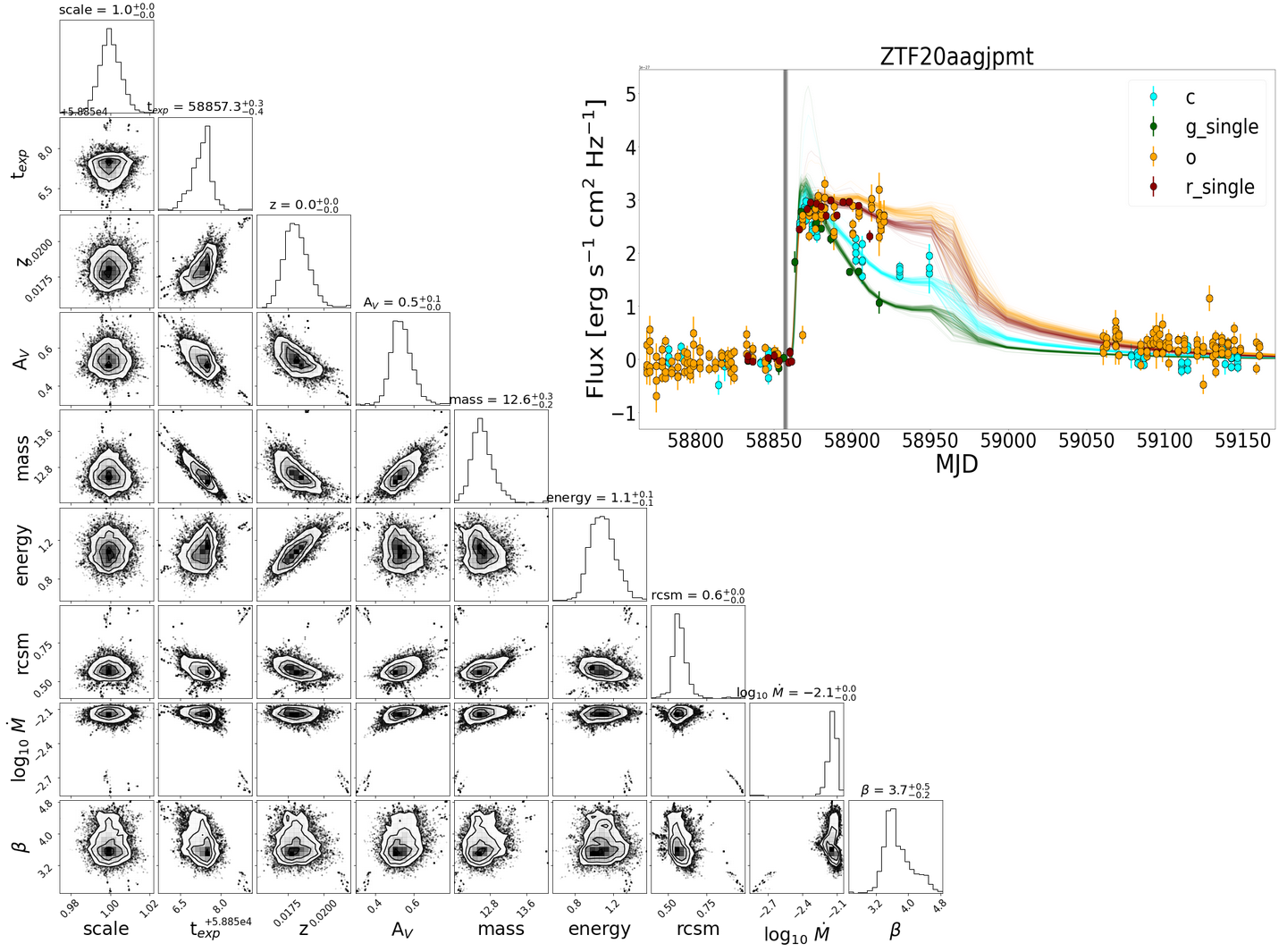}
	\caption{Same as \ref{posterior_example}, but for SN 2020aer/ZTF20aagjpmt}
	\label{posterior_fewpoints}
\end{figure}

\section{Supplement Table} \label{Supplement_table}

\startlongtable
\begin{deluxetable}{ccccccccc}

\tablecaption{SNe inferred parameters (variable z)}
\tablenum{7}
\label{tabSNe_params}

\tablehead{\colhead{ZTF oid} & \colhead{t$_{exp}$} &\colhead{Mass} & \colhead{Energy} & \colhead{$\dot{M}$} & \colhead{r$_{CSM}$} & \colhead{$\beta$} & \colhead{A$_V$} & \colhead{z}  \\
\colhead{} & \colhead{days} & \colhead{M$_\odot$} & \colhead{foe} & \colhead{M$_\odot$ yr$^{-1}$} & \colhead{10$^{15}$ cm}  & \colhead{} & \colhead{mag} & \colhead{} }  

\startdata
ZTF18aatyqds & 58243.8$_{-0.0}^{+0.0}$ & 13.23$_{-0.035}^{+0.033}$ & 1.519$_{-0.011}^{+0.011}$ & 0.009$_{-5.124}^{+1.542}$ & 0.999$_{-0.000}^{+0.000}$ & 4.998$_{-0.003}^{+0.001}$ & 0.906$_{-0.006}^{+0.006}$ & 0.024$_{-0.000}^{+0.000}$ \\
ZTF18aawyjjq & 58256.9$_{-0.4}^{+0.2}$ & 12.22$_{-0.220}^{+0.969}$ & 1.151$_{-0.033}^{+0.567}$ & 0.009$_{-0.000}^{+2.957}$ & 0.947$_{-0.207}^{+0.011}$ & 3.751$_{-0.003}^{+1.247}$ & 0.983$_{-0.148}^{+0.040}$ & 0.029$_{-0.000}^{+0.001}$ \\
ZTF18abaeqpf & 59155.0$_{-0.4}^{+0.0}$ & 12.70$_{-0.024}^{+0.955}$ & 1.251$_{-0.075}^{+0.016}$ & 7.901$_{-6.371}^{+9.139}$ & 0.498$_{-0.006}^{+0.001}$ & 3.751$_{-0.003}^{+0.597}$ & 0.000$_{-0.000}^{+0.023}$ & 0.017$_{-0.000}^{+0.000}$ \\
ZTF18abckutn & 58283.7$_{-0.1}^{+0.1}$ & 12.01$_{-0.015}^{+0.112}$ & 1.108$_{-0.033}^{+0.034}$ & 0.009$_{-0.000}^{+0.000}$ & 0.828$_{-0.036}^{+0.045}$ & 4.921$_{-0.098}^{+0.066}$ & 0.821$_{-0.033}^{+0.034}$ & 0.034$_{-0.000}^{+0.000}$ \\
ZTF18abcpmwh & 58284.6$_{-0.0}^{+0.0}$ & 12.06$_{-0.043}^{+0.056}$ & 1.928$_{-0.060}^{+0.048}$ & 0.009$_{-0.000}^{+6.163}$ & 0.876$_{-0.016}^{+0.018}$ & 4.994$_{-0.020}^{+0.005}$ & 0.512$_{-0.039}^{+0.052}$ & 0.016$_{-0.000}^{+0.000}$ \\
ZTF18abcptmt & 58284.8$_{-0.0}^{+0.0}$ & 12.00$_{-0.002}^{+0.010}$ & 1.996$_{-0.011}^{+0.002}$ & 0.009$_{-0.000}^{+0.000}$ & 0.999$_{-0.001}^{+0.000}$ & 4.780$_{-0.046}^{+0.047}$ & 0.639$_{-0.046}^{+0.056}$ & 0.045$_{-0.000}^{+0.000}$ \\
ZTF18abeajml & 58297.1$_{-0.0}^{+0.0}$ & 12.00$_{-0.001}^{+0.003}$ & 1.999$_{-0.001}^{+0.000}$ & 0.009$_{-1.111}^{+3.694}$ & 0.999$_{-0.000}^{+6.362}$ & 3.750$_{-0.000}^{+0.001}$ & 0.354$_{-0.012}^{+0.013}$ & 0.035$_{-0.000}^{+0.000}$ \\
ZTF18abgladq & 58302.0$_{-0.9}^{+0.6}$ & 12.01$_{-0.011}^{+0.061}$ & 1.042$_{-0.525}^{+0.925}$ & 0.009$_{-0.000}^{+0.000}$ & 0.999$_{-0.003}^{+0.000}$ & 3.751$_{-0.040}^{+0.037}$ & 0.317$_{-0.056}^{+0.183}$ & 0.027$_{-0.006}^{+0.006}$ \\
ZTF18abmdpwe & 58333.9$_{-0.0}^{+0.0}$ & 12.00$_{-0.001}^{+0.005}$ & 1.712$_{-0.014}^{+0.013}$ & 0.008$_{-0.000}^{+0.000}$ & 0.999$_{-0.001}^{+0.000}$ & 4.998$_{-0.004}^{+0.001}$ & 1.142$_{-0.018}^{+0.017}$ & 0.014$_{-0.000}^{+0.000}$ \\
ZTF18abokyfk & 58344.8$_{-0.0}^{+0.0}$ & 13.05$_{-0.083}^{+0.080}$ & 1.796$_{-0.017}^{+0.018}$ & 0.009$_{-0.000}^{+0.000}$ & 0.947$_{-0.010}^{+0.009}$ & 4.998$_{-0.004}^{+0.001}$ & 1.316$_{-0.011}^{+0.012}$ & 0.015$_{-0.000}^{+0.000}$ \\
ZTF18abqyvzy & 58352.9$_{-0.0}^{+0.0}$ & 12.30$_{-0.029}^{+0.029}$ & 1.928$_{-0.009}^{+0.010}$ & 0.009$_{-3.466}^{+9.419}$ & 0.999$_{-0.000}^{+5.833}$ & 4.999$_{-0.000}^{+6.408}$ & 1.018$_{-0.004}^{+0.004}$ & 0.012$_{-6.123}^{+6.466}$ \\
ZTF18absclsr & 58349.4$_{-0.0}^{+0.0}$ & 12.00$_{-0.001}^{+0.005}$ & 1.998$_{-0.004}^{+0.001}$ & 0.009$_{-5.413}^{+1.550}$ & 0.999$_{-0.000}^{+0.000}$ & 4.996$_{-0.008}^{+0.003}$ & 0.826$_{-0.013}^{+0.012}$ & 0.023$_{-0.000}^{+0.000}$ \\
ZTF18absldfl & 58357.7$_{-0.1}^{+0.1}$ & 12.00$_{-0.006}^{+0.022}$ & 1.985$_{-0.035}^{+0.013}$ & 0.007$_{-0.000}^{+0.000}$ & 0.999$_{-0.001}^{+0.000}$ & 3.749$_{-0.031}^{+0.034}$ & 0.819$_{-0.019}^{+0.020}$ & 0.032$_{-0.000}^{+0.000}$ \\
ZTF18abvmlow & 58477.8$_{-0.0}^{+0.0}$ & 12.00$_{-0.004}^{+0.017}$ & 1.625$_{-0.040}^{+0.038}$ & 0.001$_{-6.371}^{+6.697}$ & 0.999$_{-0.000}^{+0.000}$ & 4.994$_{-0.020}^{+0.004}$ & 0.274$_{-0.012}^{+0.012}$ & 0.014$_{-0.000}^{+0.000}$ \\
ZTF18acbvhit & 58402.9$_{-0.1}^{+0.1}$ & 15.41$_{-0.130}^{+0.166}$ & 1.000$_{-0.004}^{+0.004}$ & 0.000$_{-5.620}^{+6.283}$ & 0.998$_{-0.005}^{+0.001}$ & 3.749$_{-0.055}^{+0.037}$ & 0.947$_{-0.022}^{+0.021}$ & 0.014$_{-0.000}^{+0.000}$ \\
ZTF18achtnvk & 58427.7$_{-0.1}^{+0.0}$ & 12.01$_{-0.015}^{+0.053}$ & 1.996$_{-0.010}^{+0.003}$ & 0.008$_{-0.000}^{+0.001}$ & 0.966$_{-0.021}^{+0.018}$ & 4.842$_{-0.216}^{+0.133}$ & 1.182$_{-0.059}^{+0.066}$ & 0.033$_{-0.001}^{+0.001}$ \\
ZTF18acqwdla & 58439.0$_{-0.0}^{+0.0}$ & 12.73$_{-0.068}^{+0.066}$ & 1.099$_{-0.007}^{+0.008}$ & 0.009$_{-2.009}^{+5.155}$ & 0.999$_{-0.000}^{+8.130}$ & 3.750$_{-0.002}^{+0.002}$ & 0.752$_{-0.033}^{+0.022}$ & 0.016$_{-0.000}^{+0.000}$ \\
ZTF18acrtvmm & 58442.5$_{-0.7}^{+0.3}$ & 12.89$_{-0.339}^{+0.293}$ & 1.015$_{-0.050}^{+0.040}$ & 0.009$_{-0.001}^{+0.000}$ & 0.914$_{-0.028}^{+0.031}$ & 3.326$_{-0.871}^{+0.647}$ & 0.728$_{-0.108}^{+0.063}$ & 0.019$_{-0.000}^{+0.000}$ \\
ZTF18acvgyst & 58447.9$_{-0.2}^{+1.1}$ & 13.25$_{-0.530}^{+0.376}$ & 1.067$_{-0.138}^{+0.121}$ & 0.000$_{-0.000}^{+0.005}$ & 0.729$_{-0.617}^{+0.169}$ & 4.155$_{-1.648}^{+0.823}$ & 0.064$_{-0.061}^{+0.166}$ & 0.027$_{-0.001}^{+0.001}$ \\
ZTF18adbacau & 58471.3$_{-0.4}^{+1.9}$ & 14.35$_{-0.404}^{+0.163}$ & 1.728$_{-0.550}^{+0.141}$ & 0.001$_{-0.000}^{+0.006}$ & 0.488$_{-0.151}^{+0.288}$ & 4.980$_{-0.948}^{+0.018}$ & 0.297$_{-0.050}^{+0.314}$ & 0.030$_{-0.002}^{+0.001}$ \\
ZTF18adbclkd & 58470.1$_{-1.7}^{+0.5}$ & 12.64$_{-0.302}^{+1.027}$ & 0.959$_{-0.293}^{+0.038}$ & 0.007$_{-0.004}^{+0.001}$ & 0.522$_{-0.040}^{+0.258}$ & 4.722$_{-0.205}^{+0.262}$ & 0.389$_{-0.052}^{+0.293}$ & 0.015$_{-0.001}^{+0.000}$ \\
ZTF19aadnxog & 58492.4$_{-0.0}^{+0.0}$ & 15.11$_{-0.076}^{+0.077}$ & 0.697$_{-0.005}^{+0.005}$ & 0.007$_{-0.000}^{+0.000}$ & 0.999$_{-0.000}^{+8.587}$ & 4.731$_{-0.037}^{+0.036}$ & 0.707$_{-0.024}^{+0.026}$ & 0.012$_{-9.915}^{+9.571}$ \\
ZTF19aakssnv & 58514.7$_{-0.2}^{+0.3}$ & 14.08$_{-0.093}^{+0.112}$ & 1.003$_{-0.042}^{+0.033}$ & 0.003$_{-0.000}^{+0.000}$ & 0.819$_{-0.023}^{+0.025}$ & 2.870$_{-0.137}^{+0.246}$ & 0.165$_{-0.053}^{+0.060}$ & 0.025$_{-0.000}^{+0.000}$ \\
ZTF19aallimd & 58535.0$_{-0.1}^{+0.1}$ & 12.56$_{-0.096}^{+0.105}$ & 1.134$_{-0.023}^{+0.024}$ & 0.009$_{-0.000}^{+0.000}$ & 0.500$_{-0.004}^{+0.004}$ & 3.749$_{-0.040}^{+0.039}$ & 0.244$_{-0.023}^{+0.023}$ & 0.015$_{-0.000}^{+0.000}$ \\
ZTF19aalsnbp & 58534.7$_{-0.0}^{+0.2}$ & 12.90$_{-0.283}^{+0.125}$ & 1.097$_{-0.009}^{+0.019}$ & 0.009$_{-0.000}^{+3.351}$ & 0.999$_{-0.002}^{+0.000}$ & 3.749$_{-0.089}^{+0.054}$ & 0.689$_{-0.077}^{+0.061}$ & 0.037$_{-0.000}^{+0.000}$ \\
ZTF19aamggnw & 58537.6$_{-0.9}^{+0.4}$ & 12.20$_{-0.180}^{+0.259}$ & 1.082$_{-0.035}^{+0.036}$ & 0.001$_{-0.000}^{+0.008}$ & 0.970$_{-0.499}^{+0.028}$ & 3.591$_{-1.156}^{+0.407}$ & 0.017$_{-0.015}^{+0.046}$ & 0.032$_{-0.000}^{+0.000}$ \\
ZTF19aamhmsx & 58540.0$_{-0.7}^{+2.0}$ & 15.18$_{-1.054}^{+0.495}$ & 1.032$_{-0.071}^{+0.264}$ & 0.009$_{-0.004}^{+0.000}$ & 0.893$_{-0.048}^{+0.046}$ & 4.949$_{-1.147}^{+0.047}$ & 0.455$_{-0.302}^{+0.152}$ & 0.035$_{-0.001}^{+0.001}$ \\
ZTF19aamkfvy & 58549.1$_{-0.3}^{+1.9}$ & 12.80$_{-0.787}^{+0.177}$ & 1.289$_{-0.033}^{+0.306}$ & 0.006$_{-0.002}^{+0.000}$ & 0.993$_{-0.041}^{+0.005}$ & 4.944$_{-0.691}^{+0.051}$ & 0.975$_{-0.148}^{+0.052}$ & 0.026$_{-0.000}^{+0.002}$ \\
ZTF19aamljom & 58548.9$_{-0.0}^{+0.0}$ & 13.19$_{-0.028}^{+0.030}$ & 1.566$_{-0.008}^{+0.008}$ & 0.009$_{-1.390}^{+7.936}$ & 0.999$_{-0.000}^{+8.424}$ & 4.998$_{-0.004}^{+0.000}$ & 0.015$_{-0.010}^{+0.005}$ & 0.019$_{-0.000}^{+0.000}$ \\
ZTF19aamowaf & 58512.4$_{-0.2}^{+1.2}$ & 14.07$_{-0.093}^{+0.160}$ & 1.990$_{-0.102}^{+0.009}$ & 0.006$_{-0.005}^{+0.000}$ & 0.843$_{-0.028}^{+0.031}$ & 3.750$_{-0.031}^{+1.244}$ & 0.005$_{-0.004}^{+0.033}$ & 0.026$_{-0.000}^{+0.000}$ \\
ZTF19aamtwiz & 58552.4$_{-0.3}^{+0.4}$ & 12.63$_{-0.233}^{+0.348}$ & 1.158$_{-0.040}^{+0.069}$ & 0.009$_{-0.000}^{+0.000}$ & 0.799$_{-0.082}^{+0.048}$ & 4.992$_{-0.047}^{+0.007}$ & 0.264$_{-0.073}^{+0.041}$ & 0.040$_{-0.000}^{+0.001}$ \\
ZTF19aamwhat & 58543.2$_{-0.4}^{+0.3}$ & 15.91$_{-1.246}^{+0.080}$ & 0.689$_{-0.014}^{+0.335}$ & 0.003$_{-0.000}^{+0.006}$ & 0.300$_{-0.003}^{+0.140}$ & 3.749$_{-1.246}^{+0.010}$ & 0.895$_{-0.166}^{+0.026}$ & 0.011$_{-0.000}^{+0.001}$ \\
ZTF19aanhhal & 58557.1$_{-1.5}^{+0.1}$ & 12.29$_{-0.097}^{+2.298}$ & 1.149$_{-0.152}^{+0.027}$ & 0.007$_{-0.000}^{+0.000}$ & 0.504$_{-0.023}^{+0.387}$ & 4.497$_{-0.458}^{+0.191}$ & 0.276$_{-0.028}^{+0.341}$ & 0.026$_{-0.002}^{+0.000}$ \\
ZTF19aaniore & 58557.9$_{-0.2}^{+0.2}$ & 15.96$_{-0.097}^{+0.029}$ & 0.696$_{-0.009}^{+0.011}$ & 0.005$_{-0.000}^{+0.000}$ & 0.820$_{-0.024}^{+0.022}$ & 3.730$_{-0.087}^{+0.112}$ & 0.693$_{-0.055}^{+0.056}$ & 0.024$_{-0.000}^{+0.000}$ \\
ZTF19aanlekq & 58557.3$_{-0.9}^{+0.1}$ & 12.00$_{-0.002}^{+0.008}$ & 1.866$_{-0.019}^{+0.112}$ & 0.009$_{-6.306}^{+4.369}$ & 0.999$_{-0.002}^{+0.000}$ & 3.750$_{-0.002}^{+1.160}$ & 0.776$_{-0.089}^{+0.012}$ & 0.031$_{-0.000}^{+0.002}$ \\
ZTF19aanovps & 58559.7$_{-1.8}^{+0.2}$ & 13.94$_{-0.314}^{+1.542}$ & 1.116$_{-0.476}^{+0.065}$ & 0.006$_{-0.002}^{+0.000}$ & 0.746$_{-0.059}^{+0.186}$ & 3.805$_{-0.115}^{+0.680}$ & 1.007$_{-0.050}^{+0.058}$ & 0.020$_{-0.000}^{+0.000}$ \\
ZTF19aanrrqu & 58562.7$_{-0.0}^{+0.0}$ & 12.00$_{-0.001}^{+0.004}$ & 1.999$_{-0.001}^{+0.000}$ & 0.009$_{-1.532}^{+4.632}$ & 0.999$_{-0.000}^{+8.528}$ & 3.750$_{-0.002}^{+0.002}$ & 0.430$_{-0.025}^{+0.027}$ & 0.024$_{-0.000}^{+0.000}$ \\
ZTF19aapafit & 58573.6$_{-0.0}^{+0.0}$ & 12.00$_{-0.002}^{+0.007}$ & 0.923$_{-0.004}^{+0.005}$ & 0.000$_{-3.032}^{+3.756}$ & 0.944$_{-0.021}^{+0.017}$ & 3.750$_{-0.007}^{+0.008}$ & 0.017$_{-0.004}^{+0.016}$ & 0.019$_{-0.000}^{+0.000}$ \\
ZTF19aapbfot & 58570.6$_{-0.2}^{+0.2}$ & 15.33$_{-0.416}^{+0.426}$ & 1.659$_{-0.091}^{+0.067}$ & 0.009$_{-0.001}^{+0.000}$ & 0.982$_{-0.028}^{+0.015}$ & 4.966$_{-0.123}^{+0.030}$ & 0.882$_{-0.168}^{+0.047}$ & 0.036$_{-0.001}^{+0.001}$ \\
ZTF19aaqdkrm & 58579.0$_{-0.1}^{+0.2}$ & 12.21$_{-0.173}^{+0.201}$ & 1.996$_{-0.010}^{+0.003}$ & 0.009$_{-0.000}^{+0.000}$ & 0.997$_{-0.007}^{+0.001}$ & 3.752$_{-0.064}^{+0.087}$ & 0.190$_{-0.063}^{+0.049}$ & 0.039$_{-0.000}^{+0.000}$ \\
ZTF19aaqxosb & 58584.2$_{-0.1}^{+0.2}$ & 12.02$_{-0.027}^{+0.095}$ & 1.245$_{-0.040}^{+0.041}$ & 0.001$_{-0.000}^{+0.000}$ & 0.997$_{-0.006}^{+0.002}$ & 4.868$_{-1.169}^{+0.122}$ & 0.845$_{-0.034}^{+0.036}$ & 0.022$_{-0.000}^{+0.000}$ \\
ZTF19aariwfe & 58589.6$_{-0.8}^{+0.7}$ & 12.35$_{-0.312}^{+0.462}$ & 1.616$_{-0.079}^{+0.083}$ & 0.003$_{-0.001}^{+0.002}$ & 0.947$_{-0.063}^{+0.042}$ & 4.300$_{-0.786}^{+0.651}$ & 0.239$_{-0.127}^{+0.111}$ & 0.036$_{-0.002}^{+0.001}$ \\
ZTF19aarjfqe & 58588.2$_{-0.5}^{+1.0}$ & 12.94$_{-0.522}^{+1.155}$ & 1.665$_{-0.467}^{+0.166}$ & 0.007$_{-0.005}^{+0.001}$ & 0.804$_{-0.077}^{+0.192}$ & 4.972$_{-0.117}^{+0.025}$ & 0.363$_{-0.131}^{+0.173}$ & 0.028$_{-0.001}^{+0.001}$ \\
ZTF19aatlqdf & 58593.6$_{-0.3}^{+0.7}$ & 15.58$_{-0.620}^{+0.362}$ & 0.570$_{-0.048}^{+0.072}$ & 0.006$_{-0.000}^{+0.000}$ & 0.805$_{-0.051}^{+0.042}$ & 4.794$_{-0.338}^{+0.179}$ & 0.208$_{-0.088}^{+0.055}$ & 0.032$_{-0.001}^{+0.001}$ \\
ZTF19aaugaam & 58604.2$_{-0.2}^{+0.3}$ & 12.02$_{-0.021}^{+0.077}$ & 1.973$_{-0.085}^{+0.024}$ & 0.007$_{-0.006}^{+0.001}$ & 0.624$_{-0.035}^{+0.065}$ & 4.974$_{-0.108}^{+0.023}$ & 0.259$_{-0.129}^{+0.041}$ & 0.032$_{-0.000}^{+0.000}$ \\
ZTF19aauishy & 58606.8$_{-0.3}^{+0.4}$ & 12.02$_{-0.023}^{+0.108}$ & 0.963$_{-0.054}^{+0.046}$ & 0.000$_{-0.000}^{+0.000}$ & 0.991$_{-0.043}^{+0.008}$ & 1.732$_{-0.525}^{+0.262}$ & 0.255$_{-0.047}^{+0.068}$ & 0.029$_{-0.001}^{+0.001}$ \\
ZTF19aauqwna & 58608.9$_{-0.0}^{+0.0}$ & 12.02$_{-0.021}^{+0.069}$ & 1.997$_{-0.009}^{+0.002}$ & 0.009$_{-0.000}^{+8.286}$ & 0.998$_{-0.004}^{+0.001}$ & 4.990$_{-0.020}^{+0.008}$ & 1.068$_{-0.023}^{+0.023}$ & 0.031$_{-0.000}^{+0.000}$ \\
ZTF19aavbjfp & 58612.8$_{-0.1}^{+1.1}$ & 12.81$_{-0.714}^{+0.220}$ & 1.628$_{-0.049}^{+0.061}$ & 0.002$_{-0.001}^{+0.000}$ & 0.842$_{-0.056}^{+0.150}$ & 3.749$_{-0.055}^{+0.035}$ & 0.199$_{-0.093}^{+0.041}$ & 0.033$_{-0.001}^{+0.002}$ \\
ZTF19aavhblr & 58612.0$_{-0.3}^{+0.2}$ & 12.02$_{-0.022}^{+0.066}$ & 1.808$_{-0.112}^{+0.153}$ & 0.009$_{-0.000}^{+0.000}$ & 0.998$_{-0.003}^{+0.001}$ & 3.776$_{-0.137}^{+0.878}$ & 0.274$_{-0.116}^{+0.065}$ & 0.046$_{-0.002}^{+0.003}$ \\
ZTF19aavkptg & 58613.9$_{-0.8}^{+0.6}$ & 13.08$_{-0.581}^{+0.589}$ & 1.068$_{-0.215}^{+0.246}$ & 0.003$_{-0.000}^{+0.001}$ & 0.967$_{-0.057}^{+0.028}$ & 3.997$_{-1.009}^{+0.891}$ & 0.133$_{-0.084}^{+0.138}$ & 0.041$_{-0.004}^{+0.003}$ \\
ZTF19aavrcew & 58622.1$_{-0.0}^{+0.0}$ & 12.00$_{-0.002}^{+0.010}$ & 1.995$_{-0.013}^{+0.003}$ & 0.009$_{-0.000}^{+0.000}$ & 0.998$_{-0.003}^{+0.000}$ & 4.980$_{-0.047}^{+0.017}$ & 0.787$_{-0.018}^{+0.018}$ & 0.019$_{-0.000}^{+0.000}$ \\
ZTF19aawgxdn & 58628.6$_{-0.0}^{+0.0}$ & 12.00$_{-0.001}^{+0.005}$ & 0.832$_{-0.011}^{+0.011}$ & 0.007$_{-0.000}^{+0.000}$ & 0.515$_{-0.003}^{+0.004}$ & 3.750$_{-0.004}^{+0.005}$ & 0.725$_{-0.026}^{+0.028}$ & 0.022$_{-0.000}^{+0.000}$ \\
ZTF19aaycrgf & 58623.3$_{-0.7}^{+4.3}$ & 13.58$_{-0.207}^{+0.223}$ & 0.737$_{-0.079}^{+0.408}$ & 0.003$_{-0.001}^{+0.002}$ & 0.534$_{-0.081}^{+0.315}$ & 4.811$_{-0.494}^{+0.170}$ & 0.577$_{-0.063}^{+0.100}$ & 0.017$_{-0.000}^{+0.000}$ \\
ZTF19aaydtur & 58635.6$_{-0.2}^{+0.2}$ & 12.50$_{-0.293}^{+0.356}$ & 0.844$_{-0.090}^{+0.091}$ & 0.002$_{-0.000}^{+0.000}$ & 0.835$_{-0.063}^{+0.066}$ & 3.750$_{-0.159}^{+0.100}$ & 0.148$_{-0.073}^{+0.087}$ & 0.030$_{-0.001}^{+0.001}$ \\
ZTF19aazfvhh & 58640.6$_{-0.3}^{+0.3}$ & 12.02$_{-0.027}^{+0.115}$ & 1.194$_{-0.018}^{+0.021}$ & 0.006$_{-0.001}^{+0.000}$ & 0.641$_{-0.172}^{+0.118}$ & 4.152$_{-0.529}^{+0.694}$ & 0.710$_{-0.110}^{+0.095}$ & 0.033$_{-0.001}^{+0.000}$ \\
ZTF19aazyvub & 58644.3$_{-0.4}^{+0.2}$ & 12.94$_{-0.356}^{+0.324}$ & 1.311$_{-0.041}^{+0.041}$ & 0.000$_{-0.000}^{+0.000}$ & 0.998$_{-0.005}^{+0.000}$ & 4.838$_{-0.799}^{+0.150}$ & 0.003$_{-0.002}^{+0.014}$ & 0.030$_{-0.000}^{+0.000}$ \\
ZTF19abajxet & 58651.5$_{-0.8}^{+0.0}$ & 12.37$_{-0.377}^{+0.094}$ & 1.617$_{-0.085}^{+0.009}$ & 0.002$_{-9.415}^{+0.007}$ & 0.963$_{-0.151}^{+0.011}$ & 4.998$_{-0.181}^{+0.001}$ & 0.284$_{-0.013}^{+0.155}$ & 0.014$_{-0.000}^{+0.000}$ \\
ZTF19abbwfgp & 58654.6$_{-0.0}^{+0.9}$ & 13.16$_{-0.877}^{+0.075}$ & 1.582$_{-0.020}^{+0.080}$ & 0.009$_{-0.001}^{+5.290}$ & 0.961$_{-0.069}^{+0.009}$ & 4.999$_{-0.013}^{+0.000}$ & 0.425$_{-0.126}^{+0.020}$ & 0.021$_{-0.000}^{+0.000}$ \\
ZTF19abcekty & 58648.9$_{-0.0}^{+0.0}$ & 12.00$_{-0.000}^{+0.001}$ & 1.999$_{-0.000}^{+0.000}$ & 0.009$_{-1.455}^{+4.104}$ & 0.999$_{-0.000}^{+4.348}$ & 4.931$_{-0.023}^{+0.022}$ & 0.369$_{-0.012}^{+0.013}$ & 0.015$_{-0.000}^{+0.000}$ \\
ZTF19abgiwkt & 58672.2$_{-0.2}^{+0.1}$ & 13.71$_{-0.173}^{+0.663}$ & 1.136$_{-0.309}^{+0.038}$ & 0.004$_{-0.000}^{+0.001}$ & 0.751$_{-0.049}^{+0.129}$ & 3.474$_{-0.985}^{+0.283}$ & 1.235$_{-0.052}^{+0.043}$ & 0.010$_{-0.000}^{+0.000}$ \\
ZTF19abjbtbm & 58689.0$_{-0.0}^{+0.1}$ & 12.76$_{-0.218}^{+0.152}$ & 0.516$_{-0.014}^{+0.040}$ & 0.000$_{-0.000}^{+9.200}$ & 0.302$_{-0.050}^{+0.115}$ & 4.033$_{-0.713}^{+0.881}$ & 0.002$_{-0.002}^{+0.007}$ & 0.016$_{-0.000}^{+0.000}$ \\
ZTF19abjpntj & 58686.1$_{-0.6}^{+1.0}$ & 13.65$_{-0.406}^{+0.701}$ & 1.273$_{-0.095}^{+0.135}$ & 0.004$_{-0.002}^{+0.002}$ & 0.700$_{-0.095}^{+0.070}$ & 4.526$_{-1.009}^{+0.419}$ & 0.594$_{-0.457}^{+0.137}$ & 0.030$_{-0.001}^{+0.006}$ \\
ZTF19abjrjdw & 58688.5$_{-1.0}^{+0.2}$ & 12.36$_{-0.246}^{+1.241}$ & 1.585$_{-0.123}^{+0.062}$ & 0.006$_{-0.002}^{+0.002}$ & 0.888$_{-0.043}^{+0.082}$ & 4.861$_{-0.582}^{+0.129}$ & 1.108$_{-0.064}^{+0.107}$ & 0.018$_{-0.000}^{+0.000}$ \\
ZTF19abpyqog & 58705.6$_{-0.0}^{+0.0}$ & 12.26$_{-0.135}^{+0.142}$ & 1.985$_{-0.036}^{+0.013}$ & 0.006$_{-0.000}^{+0.000}$ & 0.934$_{-0.013}^{+0.013}$ & 3.751$_{-0.040}^{+0.052}$ & 0.849$_{-0.024}^{+0.026}$ & 0.037$_{-0.000}^{+0.000}$ \\
ZTF19abqgtqo & 58707.5$_{-0.2}^{+0.3}$ & 14.06$_{-0.117}^{+0.218}$ & 1.667$_{-0.064}^{+0.077}$ & 0.000$_{-0.000}^{+0.000}$ & 0.799$_{-0.044}^{+0.029}$ & 3.749$_{-0.037}^{+0.038}$ & 0.272$_{-0.027}^{+0.031}$ & 0.033$_{-0.000}^{+0.000}$ \\
ZTF19abqhobb & 58709.7$_{-0.1}^{+0.1}$ & 12.17$_{-0.123}^{+0.148}$ & 1.605$_{-0.017}^{+0.019}$ & 0.001$_{-7.041}^{+8.073}$ & 0.996$_{-0.004}^{+0.002}$ & 4.989$_{-0.022}^{+0.009}$ & 0.434$_{-0.015}^{+0.015}$ & 0.017$_{-0.000}^{+0.000}$ \\
ZTF19abqrhvt & 58706.2$_{-0.0}^{+0.1}$ & 12.00$_{-0.003}^{+0.010}$ & 1.998$_{-0.002}^{+0.000}$ & 0.009$_{-5.567}^{+2.074}$ & 0.999$_{-0.001}^{+0.000}$ & 4.993$_{-0.012}^{+0.005}$ & 0.102$_{-0.019}^{+0.025}$ & 0.025$_{-0.000}^{+0.000}$ \\
ZTF19abqrhvy & 58708.1$_{-0.0}^{+0.0}$ & 13.14$_{-0.061}^{+0.062}$ & 1.226$_{-0.022}^{+0.023}$ & 0.007$_{-0.000}^{+0.000}$ & 0.667$_{-0.021}^{+0.021}$ & 4.998$_{-0.004}^{+0.001}$ & 0.417$_{-0.020}^{+0.020}$ & 0.020$_{-0.000}^{+0.000}$ \\
ZTF19abrbmvt & 58713.3$_{-0.9}^{+0.3}$ & 12.65$_{-0.306}^{+0.762}$ & 1.375$_{-0.108}^{+0.082}$ & 0.000$_{-0.000}^{+0.000}$ & 0.970$_{-0.072}^{+0.027}$ & 4.970$_{-0.105}^{+0.027}$ & 0.324$_{-0.049}^{+0.149}$ & 0.035$_{-0.002}^{+0.001}$ \\
ZTF19abudjie & 58715.5$_{-0.5}^{+0.4}$ & 15.47$_{-0.102}^{+0.159}$ & 1.409$_{-0.156}^{+0.069}$ & 0.003$_{-0.000}^{+0.001}$ & 0.500$_{-0.009}^{+0.025}$ & 4.990$_{-0.556}^{+0.008}$ & 1.242$_{-0.067}^{+0.055}$ & 0.022$_{-0.000}^{+0.000}$ \\
ZTF19abueupg & 58716.6$_{-0.1}^{+2.6}$ & 13.94$_{-0.776}^{+0.069}$ & 1.472$_{-0.460}^{+0.037}$ & 0.009$_{-0.006}^{+7.492}$ & 0.881$_{-0.030}^{+0.097}$ & 4.796$_{-0.468}^{+0.128}$ & 0.817$_{-0.198}^{+0.031}$ & 0.030$_{-0.000}^{+0.000}$ \\
ZTF19abukbit & 58714.5$_{-0.8}^{+0.8}$ & 12.23$_{-0.213}^{+0.556}$ & 1.346$_{-0.091}^{+0.101}$ & 0.001$_{-0.000}^{+0.001}$ & 0.989$_{-0.042}^{+0.009}$ & 4.307$_{-1.314}^{+0.654}$ & 0.340$_{-0.148}^{+0.221}$ & 0.047$_{-0.003}^{+0.004}$ \\
ZTF19abwamby & 58723.5$_{-0.1}^{+0.1}$ & 12.03$_{-0.031}^{+0.104}$ & 1.888$_{-0.229}^{+0.101}$ & 0.006$_{-0.000}^{+0.000}$ & 0.998$_{-0.004}^{+0.001}$ & 3.749$_{-0.084}^{+0.084}$ & 0.293$_{-0.042}^{+0.053}$ & 0.043$_{-0.002}^{+0.001}$ \\
ZTF19abwsagv & 58726.0$_{-0.1}^{+0.0}$ & 12.00$_{-0.005}^{+0.025}$ & 1.996$_{-0.014}^{+0.003}$ & 0.007$_{-0.000}^{+0.000}$ & 0.597$_{-0.026}^{+0.034}$ & 4.971$_{-0.082}^{+0.026}$ & 0.005$_{-0.004}^{+0.016}$ & 0.041$_{-0.000}^{+0.000}$ \\
ZTF19abwztsb & 58723.8$_{-0.1}^{+0.2}$ & 14.03$_{-0.033}^{+0.058}$ & 1.994$_{-0.021}^{+0.005}$ & 0.009$_{-0.000}^{+0.000}$ & 0.738$_{-0.015}^{+0.012}$ & 4.084$_{-0.078}^{+0.089}$ & 0.824$_{-0.012}^{+0.012}$ & 0.010$_{-0.000}^{+0.000}$ \\
ZTF19abzrdup & 58736.9$_{-4.7}^{+1.2}$ & 13.43$_{-0.107}^{+0.066}$ & 1.923$_{-0.654}^{+0.069}$ & 0.004$_{-0.001}^{+0.004}$ & 0.498$_{-0.395}^{+0.009}$ & 4.918$_{-0.151}^{+0.073}$ & 0.004$_{-0.003}^{+0.010}$ & 0.037$_{-0.000}^{+0.000}$ \\
ZTF19acbhvgi & 58744.0$_{-0.6}^{+0.4}$ & 14.86$_{-0.854}^{+0.527}$ & 1.072$_{-0.037}^{+0.042}$ & 0.002$_{-0.000}^{+0.001}$ & 0.997$_{-0.014}^{+0.002}$ & 4.173$_{-0.605}^{+0.796}$ & 0.103$_{-0.062}^{+0.115}$ & 0.035$_{-0.001}^{+0.001}$ \\
ZTF19acbrzzr & 58749.5$_{-1.0}^{+0.3}$ & 15.05$_{-0.531}^{+0.617}$ & 0.971$_{-0.124}^{+0.124}$ & 0.009$_{-0.001}^{+0.000}$ & 0.850$_{-0.044}^{+0.040}$ & 2.525$_{-0.211}^{+0.798}$ & 0.557$_{-0.119}^{+0.159}$ & 0.027$_{-0.001}^{+0.001}$ \\
ZTF19acbvisk & 58746.1$_{-0.8}^{+1.0}$ & 13.24$_{-0.432}^{+0.463}$ & 1.095$_{-0.033}^{+0.036}$ & 0.007$_{-0.003}^{+0.002}$ & 0.908$_{-0.046}^{+0.037}$ & 2.769$_{-0.406}^{+1.402}$ & 0.219$_{-0.145}^{+0.150}$ & 0.035$_{-0.002}^{+0.002}$ \\
ZTF19acchaza & 58751.8$_{-0.0}^{+0.1}$ & 12.01$_{-0.010}^{+0.037}$ & 1.995$_{-0.012}^{+0.004}$ & 0.009$_{-0.000}^{+4.757}$ & 0.999$_{-0.001}^{+0.000}$ & 2.496$_{-0.757}^{+0.027}$ & 0.471$_{-0.074}^{+0.082}$ & 0.044$_{-0.001}^{+0.001}$ \\
ZTF19acewuwn & 58768.1$_{-0.5}^{+0.3}$ & 12.04$_{-0.040}^{+0.156}$ & 0.951$_{-0.049}^{+0.048}$ & 0.003$_{-0.002}^{+0.000}$ & 0.500$_{-0.019}^{+0.038}$ & 1.066$_{-0.061}^{+0.694}$ & 0.231$_{-0.052}^{+0.046}$ & 0.031$_{-0.001}^{+0.001}$ \\
ZTF19acftfav & 58767.8$_{-0.1}^{+0.1}$ & 12.03$_{-0.029}^{+0.103}$ & 1.504$_{-0.028}^{+0.034}$ & 0.009$_{-0.000}^{+0.000}$ & 0.864$_{-0.025}^{+0.023}$ & 4.996$_{-0.012}^{+0.003}$ & 1.226$_{-0.035}^{+0.033}$ & 0.015$_{-0.000}^{+0.000}$ \\
ZTF19acgbkzr & 58769.6$_{-1.3}^{+0.8}$ & 12.16$_{-0.158}^{+3.068}$ & 1.179$_{-0.253}^{+0.067}$ & 0.007$_{-0.005}^{+0.001}$ & 0.778$_{-0.129}^{+0.215}$ & 4.079$_{-1.557}^{+0.868}$ & 0.506$_{-0.173}^{+0.239}$ & 0.033$_{-0.002}^{+0.002}$ \\
ZTF19acignlo & 58775.1$_{-0.1}^{+0.1}$ & 12.00$_{-0.002}^{+0.009}$ & 0.742$_{-0.020}^{+0.020}$ & 0.007$_{-0.000}^{+0.000}$ & 0.998$_{-0.005}^{+0.001}$ & 4.711$_{-0.168}^{+0.123}$ & 0.610$_{-0.045}^{+0.047}$ & 0.023$_{-0.000}^{+0.000}$ \\
ZTF19acjwdnu & 58776.9$_{-0.2}^{+0.2}$ & 12.01$_{-0.015}^{+0.059}$ & 1.977$_{-0.060}^{+0.021}$ & 0.009$_{-0.000}^{+0.000}$ & 0.998$_{-0.005}^{+0.001}$ & 3.746$_{-0.192}^{+0.110}$ & 0.404$_{-0.075}^{+0.094}$ & 0.039$_{-0.001}^{+0.001}$ \\
ZTF19aclobbu & 58782.0$_{-0.1}^{+0.1}$ & 12.15$_{-0.070}^{+0.070}$ & 0.758$_{-0.009}^{+0.009}$ & 0.006$_{-0.000}^{+0.000}$ & 0.729$_{-0.024}^{+0.023}$ & 3.971$_{-0.092}^{+0.099}$ & 0.549$_{-0.025}^{+0.029}$ & 0.014$_{-0.000}^{+0.000}$ \\
ZTF19acryurj & 58794.8$_{-0.0}^{+0.0}$ & 12.00$_{-0.001}^{+0.006}$ & 1.999$_{-0.001}^{+0.000}$ & 0.009$_{-2.776}^{+7.949}$ & 0.999$_{-0.000}^{+0.000}$ & 3.749$_{-0.002}^{+0.003}$ & 0.584$_{-0.037}^{+0.039}$ & 0.021$_{-0.000}^{+0.000}$ \\
ZTF19acszmgx & 58791.2$_{-1.1}^{+0.7}$ & 15.26$_{-0.648}^{+0.571}$ & 1.096$_{-0.044}^{+0.332}$ & 0.003$_{-0.001}^{+0.005}$ & 0.994$_{-0.107}^{+0.004}$ & 4.735$_{-0.798}^{+0.236}$ & 0.380$_{-0.283}^{+0.153}$ & 0.034$_{-0.001}^{+0.003}$ \\
ZTF19acvrjzd & 58801.2$_{-0.3}^{+1.1}$ & 13.13$_{-1.099}^{+0.292}$ & 1.995$_{-0.017}^{+0.004}$ & 0.009$_{-0.006}^{+0.000}$ & 0.993$_{-0.011}^{+0.005}$ & 2.510$_{-0.325}^{+2.485}$ & 0.106$_{-0.051}^{+0.084}$ & 0.041$_{-0.001}^{+0.001}$ \\
ZTF19acwrrvg & 58809.3$_{-0.0}^{+0.0}$ & 12.00$_{-0.000}^{+0.003}$ & 1.999$_{-0.001}^{+0.000}$ & 0.009$_{-9.646}^{+2.930}$ & 0.999$_{-0.000}^{+7.072}$ & 3.750$_{-0.000}^{+0.000}$ & 0.112$_{-0.017}^{+0.019}$ & 0.027$_{-0.000}^{+0.000}$ \\
ZTF19acxowrr & 58817.5$_{-0.0}^{+1.1}$ & 15.98$_{-0.041}^{+0.012}$ & 1.084$_{-0.224}^{+0.040}$ & 0.009$_{-0.000}^{+3.234}$ & 0.889$_{-0.077}^{+0.027}$ & 4.994$_{-1.248}^{+0.005}$ & 0.003$_{-0.002}^{+0.004}$ & 0.041$_{-0.005}^{+0.000}$ \\
ZTF19acyplkt & 58825.2$_{-0.3}^{+0.3}$ & 12.81$_{-0.348}^{+0.260}$ & 1.996$_{-0.009}^{+0.002}$ & 0.009$_{-8.138}^{+2.024}$ & 0.999$_{-0.001}^{+0.000}$ & 2.502$_{-0.061}^{+0.310}$ & 0.262$_{-0.033}^{+0.035}$ & 0.034$_{-0.000}^{+0.000}$ \\
ZTF19acytcsg & 58827.7$_{-0.2}^{+1.0}$ & 12.77$_{-0.179}^{+0.153}$ & 1.233$_{-0.136}^{+0.043}$ & 0.002$_{-0.002}^{+0.000}$ & 0.678$_{-0.065}^{+0.317}$ & 3.757$_{-0.032}^{+1.238}$ & 0.002$_{-0.002}^{+0.006}$ & 0.028$_{-0.000}^{+0.000}$ \\
ZTF19adbryab & 58834.6$_{-0.6}^{+1.2}$ & 12.05$_{-0.048}^{+0.437}$ & 0.555$_{-0.050}^{+0.123}$ & 0.000$_{-0.000}^{+0.004}$ & 0.960$_{-0.444}^{+0.035}$ & 1.853$_{-0.317}^{+1.134}$ & 0.980$_{-0.099}^{+0.267}$ & 0.021$_{-0.001}^{+0.001}$ \\
ZTF20aadchdd & 58840.5$_{-0.3}^{+0.1}$ & 12.00$_{-0.006}^{+0.019}$ & 0.517$_{-0.013}^{+0.014}$ & 0.009$_{-6.474}^{+2.095}$ & 0.999$_{-0.001}^{+0.000}$ & 3.749$_{-0.020}^{+0.019}$ & 0.217$_{-0.066}^{+0.044}$ & 0.036$_{-0.001}^{+0.000}$ \\
ZTF20aaekbdr & 58847.4$_{-0.3}^{+0.2}$ & 12.03$_{-0.028}^{+0.103}$ & 1.153$_{-0.060}^{+0.066}$ & 0.009$_{-0.000}^{+0.000}$ & 0.997$_{-0.008}^{+0.002}$ & 3.751$_{-0.057}^{+0.071}$ & 1.125$_{-0.087}^{+0.097}$ & 0.036$_{-0.001}^{+0.002}$ \\
ZTF20aafclxb & 58851.5$_{-0.1}^{+0.3}$ & 12.00$_{-0.005}^{+0.032}$ & 1.819$_{-0.015}^{+0.018}$ & 0.003$_{-9.861}^{+0.000}$ & 0.998$_{-0.004}^{+0.001}$ & 4.985$_{-1.048}^{+0.013}$ & 0.746$_{-0.064}^{+0.031}$ & 0.015$_{-0.000}^{+0.000}$ \\
ZTF20aagjpmt & 58857.3$_{-0.7}^{+0.4}$ & 12.60$_{-0.326}^{+0.491}$ & 1.079$_{-0.184}^{+0.237}$ & 0.007$_{-0.001}^{+0.000}$ & 0.571$_{-0.055}^{+0.085}$ & 3.651$_{-0.369}^{+0.783}$ & 0.533$_{-0.081}^{+0.104}$ & 0.018$_{-0.001}^{+0.001}$ \\
ZTF20aagnbes & 58858.1$_{-0.1}^{+0.1}$ & 12.74$_{-0.150}^{+0.121}$ & 1.022$_{-0.010}^{+0.009}$ & 0.009$_{-0.000}^{+4.472}$ & 0.999$_{-0.001}^{+0.000}$ & 3.750$_{-0.015}^{+0.026}$ & 1.007$_{-0.029}^{+0.025}$ & 0.019$_{-0.000}^{+0.000}$ \\
ZTF20aahqbsr & 58860.2$_{-1.2}^{+1.0}$ & 14.09$_{-0.132}^{+0.686}$ & 1.107$_{-0.231}^{+0.084}$ & 0.004$_{-0.001}^{+0.003}$ & 0.710$_{-0.199}^{+0.045}$ & 4.467$_{-3.102}^{+0.404}$ & 0.995$_{-0.196}^{+0.140}$ & 0.020$_{-0.000}^{+0.002}$ \\
ZTF20aahqbun & 58856.5$_{-0.3}^{+0.4}$ & 13.62$_{-0.263}^{+0.253}$ & 1.297$_{-0.042}^{+0.051}$ & 0.009$_{-0.000}^{+0.000}$ & 0.999$_{-0.001}^{+0.000}$ & 4.986$_{-0.043}^{+0.012}$ & 1.259$_{-0.056}^{+0.058}$ & 0.024$_{-0.000}^{+0.000}$ \\
ZTF20aanvqbi & 58889.5$_{-0.0}^{+0.0}$ & 13.49$_{-0.748}^{+0.119}$ & 1.949$_{-0.779}^{+0.048}$ & 0.009$_{-0.000}^{+9.158}$ & 0.674$_{-0.048}^{+0.158}$ & 3.752$_{-0.015}^{+1.230}$ & 0.708$_{-0.094}^{+0.091}$ & 0.023$_{-0.000}^{+0.001}$ \\
ZTF20aaoldej & 58889.6$_{-0.2}^{+0.2}$ & 12.21$_{-0.190}^{+0.246}$ & 1.541$_{-0.030}^{+0.032}$ & 0.005$_{-0.000}^{+0.000}$ & 0.994$_{-0.013}^{+0.005}$ & 3.970$_{-0.342}^{+0.791}$ & 0.688$_{-0.126}^{+0.102}$ & 0.031$_{-0.000}^{+0.001}$ \\
ZTF20aapycrh & 58897.1$_{-2.4}^{+0.0}$ & 12.27$_{-0.151}^{+3.284}$ & 1.536$_{-0.361}^{+0.027}$ & 0.000$_{-8.475}^{+0.002}$ & 0.956$_{-0.095}^{+0.032}$ & 3.822$_{-0.742}^{+0.783}$ & 0.002$_{-0.002}^{+0.052}$ & 0.037$_{-0.000}^{+0.000}$ \\
ZTF20aascvvo & 58898.9$_{-0.1}^{+0.1}$ & 15.99$_{-0.007}^{+0.002}$ & 1.068$_{-0.007}^{+0.007}$ & 0.009$_{-9.373}^{+2.730}$ & 0.999$_{-0.000}^{+4.038}$ & 4.856$_{-0.018}^{+0.016}$ & 0.693$_{-0.024}^{+0.027}$ & 0.017$_{-0.000}^{+0.000}$ \\
ZTF20aasjzhg & 58903.9$_{-0.2}^{+1.5}$ & 13.91$_{-0.872}^{+0.095}$ & 0.992$_{-0.030}^{+0.014}$ & 0.000$_{-6.625}^{+0.009}$ & 0.994$_{-0.159}^{+0.005}$ & 4.903$_{-1.162}^{+0.092}$ & 0.045$_{-0.043}^{+0.127}$ & 0.014$_{-0.000}^{+0.000}$ \\
ZTF20aatqgeo & 58919.2$_{-0.0}^{+0.0}$ & 14.88$_{-0.282}^{+0.288}$ & 1.986$_{-0.038}^{+0.012}$ & 0.009$_{-0.000}^{+5.675}$ & 0.999$_{-0.001}^{+0.000}$ & 3.765$_{-0.097}^{+0.072}$ & 0.555$_{-0.060}^{+0.067}$ & 0.038$_{-0.000}^{+0.000}$ \\
ZTF20aatqidk & 58919.0$_{-0.8}^{+1.9}$ & 13.27$_{-0.222}^{+0.217}$ & 1.308$_{-0.090}^{+0.396}$ & 0.007$_{-0.001}^{+0.002}$ & 0.647$_{-0.094}^{+0.077}$ & 4.966$_{-0.207}^{+0.030}$ & 0.437$_{-0.128}^{+0.230}$ & 0.032$_{-0.001}^{+0.001}$ \\
ZTF20aattqle & 58917.9$_{-0.3}^{+0.2}$ & 13.86$_{-0.194}^{+0.289}$ & 1.983$_{-0.041}^{+0.015}$ & 0.006$_{-0.000}^{+0.000}$ & 0.999$_{-0.001}^{+0.000}$ & 4.739$_{-0.103}^{+0.093}$ & 0.647$_{-0.085}^{+0.092}$ & 0.028$_{-0.000}^{+0.000}$ \\
ZTF20aatwisu & 58916.0$_{-0.3}^{+1.7}$ & 12.00$_{-0.007}^{+0.035}$ & 1.476$_{-0.808}^{+0.076}$ & 0.009$_{-0.000}^{+6.448}$ & 0.998$_{-0.003}^{+0.001}$ & 4.982$_{-1.234}^{+0.016}$ & 0.235$_{-0.051}^{+0.074}$ & 0.031$_{-0.008}^{+0.001}$ \\
ZTF20aaunfpj & 58929.6$_{-4.1}^{+0.1}$ & 14.13$_{-0.115}^{+1.840}$ & 1.995$_{-0.515}^{+0.004}$ & 0.009$_{-0.004}^{+5.511}$ & 0.999$_{-0.003}^{+0.000}$ & 4.998$_{-0.298}^{+0.000}$ & 1.779$_{-0.299}^{+0.044}$ & 0.009$_{-0.000}^{+0.000}$ \\
ZTF20aauoipy & 58918.3$_{-0.4}^{+0.6}$ & 15.92$_{-0.233}^{+0.073}$ & 1.554$_{-0.060}^{+0.104}$ & 0.009$_{-0.005}^{+0.000}$ & 0.971$_{-0.059}^{+0.025}$ & 3.766$_{-0.086}^{+1.218}$ & 0.618$_{-0.168}^{+0.143}$ & 0.035$_{-0.001}^{+0.001}$ \\
ZTF20aauoktk & 58933.3$_{-0.4}^{+0.5}$ & 14.96$_{-1.207}^{+0.764}$ & 0.623$_{-0.057}^{+0.103}$ & 0.006$_{-0.000}^{+0.001}$ & 0.997$_{-0.029}^{+0.002}$ & 4.861$_{-0.434}^{+0.127}$ & 0.004$_{-0.003}^{+0.013}$ & 0.040$_{-0.001}^{+0.001}$ \\
ZTF20aauompx & 58925.9$_{-0.0}^{+0.1}$ & 14.72$_{-0.439}^{+0.104}$ & 1.403$_{-0.064}^{+0.254}$ & 0.002$_{-0.000}^{+0.000}$ & 0.577$_{-0.037}^{+0.157}$ & 1.208$_{-0.146}^{+0.271}$ & 1.078$_{-0.052}^{+0.068}$ & 0.020$_{-0.000}^{+0.000}$ \\
ZTF20aauqhka & 58935.6$_{-0.2}^{+0.1}$ & 15.96$_{-0.104}^{+0.031}$ & 1.308$_{-0.045}^{+0.043}$ & 0.009$_{-7.222}^{+2.066}$ & 0.991$_{-0.009}^{+0.006}$ & 4.998$_{-0.004}^{+0.001}$ & 0.001$_{-0.001}^{+0.003}$ & 0.038$_{-0.000}^{+0.000}$ \\
ZTF20aauqlwv & 58924.9$_{-0.1}^{+0.1}$ & 12.00$_{-0.007}^{+0.025}$ & 1.999$_{-0.001}^{+0.000}$ & 0.009$_{-0.000}^{+9.765}$ & 0.998$_{-0.005}^{+0.001}$ & 3.749$_{-0.024}^{+0.020}$ & 0.662$_{-0.105}^{+0.100}$ & 0.029$_{-0.000}^{+0.000}$ \\
ZTF20aaurjbj & 58946.3$_{-0.0}^{+0.0}$ & 12.00$_{-0.007}^{+0.023}$ & 1.994$_{-0.014}^{+0.004}$ & 0.009$_{-5.236}^{+1.770}$ & 0.999$_{-0.000}^{+0.000}$ & 4.873$_{-0.029}^{+0.025}$ & 0.001$_{-0.001}^{+0.002}$ & 0.039$_{-0.000}^{+0.000}$ \\
ZTF20aausahr & 58944.4$_{-0.9}^{+1.0}$ & 12.33$_{-0.310}^{+0.946}$ & 1.483$_{-0.564}^{+0.447}$ & 0.006$_{-0.002}^{+0.002}$ & 0.915$_{-0.072}^{+0.075}$ & 4.829$_{-1.174}^{+0.161}$ & 0.651$_{-0.203}^{+0.167}$ & 0.033$_{-0.005}^{+0.004}$ \\
ZTF20aauvjws & 58939.5$_{-0.2}^{+0.3}$ & 12.02$_{-0.018}^{+0.059}$ & 1.926$_{-0.049}^{+0.049}$ & 0.008$_{-0.000}^{+0.000}$ & 0.990$_{-0.016}^{+0.008}$ & 4.987$_{-0.042}^{+0.011}$ & 0.984$_{-0.037}^{+0.037}$ & 0.026$_{-0.000}^{+0.000}$ \\
ZTF20aavdcxo & 58947.1$_{-0.5}^{+0.3}$ & 12.02$_{-0.024}^{+0.097}$ & 1.992$_{-0.025}^{+0.007}$ & 0.009$_{-0.000}^{+0.000}$ & 0.984$_{-0.025}^{+0.013}$ & 4.904$_{-0.079}^{+0.072}$ & 1.079$_{-0.061}^{+0.057}$ & 0.038$_{-0.001}^{+0.001}$ \\
ZTF20aavptjf & 58953.2$_{-0.3}^{+0.2}$ & 12.02$_{-0.021}^{+0.111}$ & 1.833$_{-0.182}^{+0.136}$ & 0.009$_{-0.000}^{+0.000}$ & 0.967$_{-0.014}^{+0.014}$ & 4.991$_{-0.029}^{+0.007}$ & 0.719$_{-0.088}^{+0.067}$ & 0.030$_{-0.001}^{+0.001}$ \\
ZTF20aaynrrh & 58968.9$_{-0.0}^{+0.0}$ & 12.00$_{-0.000}^{+0.001}$ & 1.999$_{-0.000}^{+0.000}$ & 0.009$_{-3.933}^{+1.805}$ & 0.999$_{-0.000}^{+7.199}$ & 4.998$_{-0.004}^{+0.001}$ & 0.376$_{-0.006}^{+0.006}$ & 0.007$_{-4.034}^{+4.030}$ \\
ZTF20aazcnrv & 58967.0$_{-0.6}^{+0.8}$ & 14.79$_{-1.340}^{+0.553}$ & 1.489$_{-0.066}^{+0.146}$ & 0.008$_{-0.000}^{+0.000}$ & 0.996$_{-0.113}^{+0.003}$ & 4.958$_{-0.104}^{+0.038}$ & 0.933$_{-0.083}^{+0.046}$ & 0.024$_{-0.001}^{+0.000}$ \\
ZTF20aazpphd & 58975.8$_{-0.2}^{+1.9}$ & 12.62$_{-0.579}^{+0.333}$ & 1.171$_{-0.060}^{+0.069}$ & 0.009$_{-0.007}^{+0.000}$ & 0.767$_{-0.147}^{+0.226}$ & 3.755$_{-0.142}^{+1.219}$ & 0.776$_{-0.293}^{+0.061}$ & 0.033$_{-0.001}^{+0.003}$ \\
ZTF20aazrxef & 58977.7$_{-0.1}^{+0.1}$ & 12.00$_{-0.008}^{+0.029}$ & 1.735$_{-0.166}^{+0.155}$ & 0.009$_{-0.000}^{+0.000}$ & 0.817$_{-0.034}^{+0.031}$ & 4.983$_{-0.053}^{+0.015}$ & 0.380$_{-0.049}^{+0.053}$ & 0.030$_{-0.001}^{+0.001}$ \\
ZTF20aazswwk & 58978.9$_{-0.2}^{+0.2}$ & 12.11$_{-0.103}^{+0.211}$ & 1.171$_{-0.035}^{+0.037}$ & 0.000$_{-6.780}^{+8.216}$ & 0.995$_{-0.014}^{+0.003}$ & 4.207$_{-0.792}^{+0.414}$ & 0.045$_{-0.027}^{+0.036}$ & 0.041$_{-0.000}^{+0.000}$ \\
ZTF20abccixp & 58990.8$_{-0.0}^{+0.0}$ & 12.00$_{-0.003}^{+0.009}$ & 1.998$_{-0.003}^{+0.000}$ & 0.009$_{-3.683}^{+1.228}$ & 0.999$_{-0.000}^{+0.000}$ & 3.750$_{-0.003}^{+0.003}$ & 0.243$_{-0.015}^{+0.016}$ & 0.037$_{-0.000}^{+0.000}$ \\
ZTF20abfcdkj & 59005.5$_{-0.6}^{+0.7}$ & 13.83$_{-0.557}^{+0.382}$ & 0.706$_{-0.060}^{+0.115}$ & 0.005$_{-0.001}^{+0.004}$ & 0.965$_{-0.184}^{+0.032}$ & 4.973$_{-0.147}^{+0.025}$ & 0.735$_{-0.107}^{+0.066}$ & 0.035$_{-0.001}^{+0.001}$ \\
ZTF20abjaapj & 59020.7$_{-0.2}^{+0.2}$ & 12.69$_{-0.086}^{+0.080}$ & 1.899$_{-0.047}^{+0.053}$ & 0.008$_{-0.000}^{+0.000}$ & 0.661$_{-0.028}^{+0.029}$ & 4.995$_{-0.016}^{+0.004}$ & 0.768$_{-0.050}^{+0.028}$ & 0.025$_{-0.000}^{+0.000}$ \\
ZTF20abjatqy & 59022.1$_{-0.1}^{+0.7}$ & 13.93$_{-0.354}^{+0.135}$ & 1.865$_{-0.080}^{+0.097}$ & 0.006$_{-0.004}^{+0.000}$ & 0.689$_{-0.059}^{+0.018}$ & 3.753$_{-0.039}^{+1.224}$ & 1.275$_{-0.078}^{+0.042}$ & 0.014$_{-0.000}^{+0.000}$ \\
ZTF20abjcyhg & 59024.1$_{-0.1}^{+0.0}$ & 12.00$_{-0.004}^{+0.018}$ & 1.928$_{-0.074}^{+0.059}$ & 0.009$_{-0.000}^{+0.000}$ & 0.999$_{-0.001}^{+0.000}$ & 4.729$_{-0.092}^{+0.094}$ & 0.477$_{-0.039}^{+0.040}$ & 0.025$_{-0.000}^{+0.000}$ \\
ZTF20abjonjs & 59027.8$_{-0.1}^{+0.1}$ & 13.41$_{-0.149}^{+0.139}$ & 0.758$_{-0.015}^{+0.016}$ & 0.002$_{-0.000}^{+0.000}$ & 0.997$_{-0.010}^{+0.002}$ & 3.748$_{-0.085}^{+0.080}$ & 0.746$_{-0.080}^{+0.046}$ & 0.018$_{-0.000}^{+0.000}$ \\
ZTF20abjyorg & 59027.5$_{-0.8}^{+1.0}$ & 13.81$_{-0.597}^{+0.976}$ & 1.988$_{-0.039}^{+0.010}$ & 0.009$_{-0.001}^{+0.000}$ & 0.998$_{-0.005}^{+0.001}$ & 3.375$_{-0.936}^{+1.057}$ & 0.272$_{-0.074}^{+0.091}$ & 0.057$_{-0.002}^{+0.002}$ \\
ZTF20ablklei & 59037.2$_{-0.1}^{+0.2}$ & 12.31$_{-0.230}^{+0.103}$ & 1.757$_{-0.026}^{+0.031}$ & 0.008$_{-0.000}^{+0.000}$ & 0.995$_{-0.018}^{+0.003}$ & 4.997$_{-0.007}^{+0.001}$ & 0.602$_{-0.043}^{+0.039}$ & 0.026$_{-0.000}^{+0.000}$ \\
ZTF20abqferm & 59060.5$_{-0.2}^{+0.2}$ & 12.05$_{-0.049}^{+0.142}$ & 1.718$_{-0.059}^{+0.063}$ & 0.002$_{-0.000}^{+0.000}$ & 0.995$_{-0.015}^{+0.004}$ & 4.464$_{-0.994}^{+0.399}$ & 0.258$_{-0.066}^{+0.057}$ & 0.045$_{-0.001}^{+0.001}$ \\
ZTF20abupxie & 59072.4$_{-0.0}^{+0.1}$ & 12.01$_{-0.011}^{+0.045}$ & 1.107$_{-0.013}^{+0.011}$ & 0.008$_{-0.000}^{+0.000}$ & 0.999$_{-0.001}^{+0.000}$ & 4.994$_{-0.025}^{+0.004}$ & 0.898$_{-0.019}^{+0.024}$ & 0.015$_{-0.000}^{+0.000}$ \\
ZTF20abuqali & 59063.1$_{-0.1}^{+0.5}$ & 15.87$_{-0.717}^{+0.114}$ & 1.109$_{-0.059}^{+0.313}$ & 0.009$_{-0.002}^{+0.000}$ & 0.993$_{-0.019}^{+0.005}$ & 3.751$_{-0.079}^{+1.199}$ & 1.012$_{-0.349}^{+0.083}$ & 0.032$_{-0.001}^{+0.006}$ \\
ZTF20abwdaeo & 59074.9$_{-0.3}^{+0.3}$ & 13.15$_{-0.153}^{+0.151}$ & 0.908$_{-0.035}^{+0.039}$ & 0.004$_{-0.000}^{+0.000}$ & 0.996$_{-0.010}^{+0.002}$ & 4.950$_{-0.110}^{+0.045}$ & 0.799$_{-0.056}^{+0.033}$ & 0.020$_{-0.000}^{+0.000}$ \\
ZTF20abybeex & 59085.2$_{-0.2}^{+0.6}$ & 15.96$_{-0.949}^{+0.030}$ & 0.616$_{-0.050}^{+0.414}$ & 0.009$_{-0.000}^{+0.000}$ & 0.500$_{-0.005}^{+0.006}$ & 3.740$_{-0.301}^{+0.392}$ & 0.711$_{-0.098}^{+0.062}$ & 0.022$_{-0.000}^{+0.001}$ \\
ZTF20abywoaa & 59086.2$_{-0.2}^{+0.2}$ & 12.02$_{-0.022}^{+0.075}$ & 1.987$_{-0.034}^{+0.011}$ & 0.009$_{-0.000}^{+8.930}$ & 0.997$_{-0.007}^{+0.002}$ & 4.988$_{-0.036}^{+0.010}$ & 0.466$_{-0.072}^{+0.109}$ & 0.046$_{-0.001}^{+0.001}$ \\
ZTF20abywydb & 59091.0$_{-0.2}^{+0.2}$ & 12.00$_{-0.006}^{+0.022}$ & 1.992$_{-0.021}^{+0.006}$ & 0.009$_{-7.608}^{+2.381}$ & 0.999$_{-0.001}^{+0.000}$ & 3.750$_{-0.019}^{+0.023}$ & 0.271$_{-0.041}^{+0.043}$ & 0.054$_{-0.001}^{+0.001}$ \\
ZTF20abyylgi & 59098.0$_{-0.0}^{+0.0}$ & 13.44$_{-0.030}^{+0.033}$ & 1.998$_{-0.005}^{+0.001}$ & 0.007$_{-0.000}^{+0.000}$ & 0.574$_{-0.011}^{+0.012}$ & 4.999$_{-0.002}^{+0.000}$ & 0.000$_{-0.000}^{+0.001}$ & 0.020$_{-0.000}^{+0.000}$ \\
ZTF20abyzomt & 59097.6$_{-0.0}^{+0.0}$ & 12.00$_{-0.004}^{+0.017}$ & 1.590$_{-0.018}^{+0.018}$ & 0.007$_{-0.000}^{+0.000}$ & 0.990$_{-0.006}^{+0.006}$ & 4.997$_{-0.008}^{+0.002}$ & 0.707$_{-0.053}^{+0.051}$ & 0.021$_{-0.000}^{+0.000}$ \\
ZTF20abyzprl & 59097.0$_{-0.2}^{+0.2}$ & 14.16$_{-0.133}^{+0.127}$ & 1.283$_{-0.077}^{+0.095}$ & 0.004$_{-0.000}^{+0.000}$ & 0.704$_{-0.038}^{+0.034}$ & 4.942$_{-0.114}^{+0.051}$ & 0.734$_{-0.085}^{+0.050}$ & 0.024$_{-0.000}^{+0.000}$ \\
ZTF20abzoaas & 59096.8$_{-0.2}^{+0.1}$ & 12.00$_{-0.001}^{+0.005}$ & 0.568$_{-0.025}^{+0.022}$ & 0.009$_{-1.833}^{+4.823}$ & 0.999$_{-0.000}^{+0.000}$ & 3.750$_{-0.002}^{+0.002}$ & 0.468$_{-0.033}^{+0.033}$ & 0.022$_{-0.000}^{+0.000}$ \\
ZTF20abzxihn & 59097.8$_{-0.6}^{+1.2}$ & 12.03$_{-0.027}^{+0.112}$ & 1.957$_{-0.191}^{+0.039}$ & 0.009$_{-0.000}^{+8.233}$ & 0.997$_{-0.007}^{+0.002}$ & 3.746$_{-1.252}^{+0.054}$ & 0.305$_{-0.112}^{+0.105}$ & 0.090$_{-0.008}^{+0.008}$ \\
ZTF20accrldu & 59108.6$_{-1.0}^{+0.1}$ & 12.27$_{-0.250}^{+0.952}$ & 1.710$_{-0.099}^{+0.065}$ & 0.008$_{-0.001}^{+0.001}$ & 0.969$_{-0.022}^{+0.019}$ & 3.749$_{-0.215}^{+0.123}$ & 0.725$_{-0.120}^{+0.081}$ & 0.037$_{-0.001}^{+0.001}$ \\
ZTF20acectxy & 59110.6$_{-0.1}^{+0.1}$ & 12.02$_{-0.020}^{+0.069}$ & 1.691$_{-0.166}^{+0.171}$ & 0.009$_{-0.000}^{+0.000}$ & 0.998$_{-0.006}^{+0.001}$ & 4.990$_{-0.030}^{+0.009}$ & 0.416$_{-0.069}^{+0.080}$ & 0.044$_{-0.002}^{+0.002}$ \\
ZTF20acfdmex & 59114.3$_{-0.0}^{+0.0}$ & 12.85$_{-0.054}^{+0.074}$ & 1.000$_{-0.001}^{+0.003}$ & 0.002$_{-6.541}^{+2.242}$ & 0.740$_{-0.026}^{+0.026}$ & 3.749$_{-0.004}^{+0.004}$ & 0.317$_{-0.022}^{+0.025}$ & 0.021$_{-0.000}^{+0.000}$ \\
ZTF20acfkyll & 59112.9$_{-0.1}^{+1.6}$ & 13.83$_{-0.129}^{+0.138}$ & 1.068$_{-0.340}^{+0.046}$ & 0.009$_{-0.006}^{+0.000}$ & 0.714$_{-0.035}^{+0.285}$ & 3.739$_{-0.809}^{+0.020}$ & 0.441$_{-0.085}^{+0.057}$ & 0.018$_{-0.000}^{+0.000}$ \\
ZTF20acfkzcg & 59115.4$_{-0.7}^{+0.0}$ & 12.12$_{-0.120}^{+0.588}$ & 0.937$_{-0.027}^{+0.061}$ & 0.003$_{-0.000}^{+0.000}$ & 0.500$_{-0.003}^{+0.013}$ & 3.765$_{-0.022}^{+0.981}$ & 0.414$_{-0.066}^{+0.036}$ & 0.019$_{-0.000}^{+0.000}$ \\
ZTF20acfvgdp & 59114.7$_{-0.3}^{+0.2}$ & 12.01$_{-0.013}^{+0.060}$ & 1.000$_{-0.015}^{+0.016}$ & 0.001$_{-0.000}^{+0.004}$ & 0.500$_{-0.024}^{+0.038}$ & 3.143$_{-1.412}^{+0.709}$ & 0.134$_{-0.060}^{+0.081}$ & 0.030$_{-0.000}^{+0.000}$ \\
ZTF20acgided & 59116.4$_{-0.0}^{+0.1}$ & 13.10$_{-0.062}^{+0.065}$ & 1.998$_{-0.004}^{+0.001}$ & 0.008$_{-0.000}^{+0.000}$ & 0.555$_{-0.023}^{+0.028}$ & 4.997$_{-0.007}^{+0.002}$ & 0.023$_{-0.010}^{+0.025}$ & 0.021$_{-0.000}^{+0.000}$ \\
ZTF20acgnelh & 59116.3$_{-1.0}^{+1.2}$ & 13.60$_{-0.775}^{+1.369}$ & 1.150$_{-0.103}^{+0.091}$ & 0.005$_{-0.003}^{+0.003}$ & 0.981$_{-0.127}^{+0.016}$ & 4.496$_{-0.964}^{+0.462}$ & 0.332$_{-0.163}^{+0.246}$ & 0.034$_{-0.002}^{+0.002}$ \\
ZTF20acgoxns & 59135.5$_{-0.0}^{+0.1}$ & 12.03$_{-0.029}^{+3.583}$ & 1.990$_{-0.092}^{+0.008}$ & 0.002$_{-0.001}^{+0.000}$ & 0.996$_{-0.611}^{+0.002}$ & 3.774$_{-0.083}^{+1.100}$ & 1.462$_{-1.348}^{+0.147}$ & 0.020$_{-0.001}^{+0.017}$ \\
ZTF20achbejn & 59119.6$_{-0.3}^{+0.4}$ & 13.25$_{-0.462}^{+0.398}$ & 1.137$_{-0.046}^{+0.053}$ & 0.006$_{-0.001}^{+0.001}$ & 0.988$_{-0.043}^{+0.010}$ & 4.943$_{-0.347}^{+0.052}$ & 0.806$_{-0.125}^{+0.074}$ & 0.033$_{-0.001}^{+0.001}$ \\
ZTF20achuhlt & 59114.3$_{-0.3}^{+0.3}$ & 14.12$_{-0.160}^{+0.416}$ & 1.984$_{-0.051}^{+0.014}$ & 0.001$_{-0.000}^{+0.000}$ & 0.706$_{-0.371}^{+0.128}$ & 1.115$_{-0.068}^{+0.120}$ & 0.002$_{-0.002}^{+0.009}$ & 0.052$_{-0.001}^{+0.001}$ \\
ZTF20acitoie & 59127.8$_{-0.3}^{+0.8}$ & 12.01$_{-0.009}^{+0.040}$ & 1.052$_{-0.034}^{+0.033}$ & 0.004$_{-0.001}^{+0.005}$ & 0.556$_{-0.168}^{+0.070}$ & 3.737$_{-1.280}^{+0.104}$ & 0.645$_{-0.090}^{+0.123}$ & 0.022$_{-0.000}^{+0.000}$ \\
ZTF20aciubfx & 59124.4$_{-0.1}^{+0.5}$ & 12.00$_{-0.005}^{+0.015}$ & 1.997$_{-0.007}^{+0.002}$ & 0.009$_{-5.398}^{+1.479}$ & 0.999$_{-0.000}^{+0.000}$ & 3.750$_{-0.013}^{+0.013}$ & 1.148$_{-0.044}^{+0.044}$ & 0.027$_{-0.000}^{+0.000}$ \\
ZTF20acknpig & 59134.3$_{-0.2}^{+0.1}$ & 12.74$_{-0.168}^{+0.289}$ & 0.926$_{-0.044}^{+0.113}$ & 0.002$_{-0.000}^{+0.000}$ & 0.500$_{-0.009}^{+0.012}$ & 3.749$_{-0.029}^{+0.025}$ & 0.435$_{-0.062}^{+0.069}$ & 0.023$_{-0.000}^{+0.000}$ \\
ZTF20acmaaan & 59138.1$_{-0.1}^{+0.1}$ & 12.04$_{-0.042}^{+0.127}$ & 1.396$_{-0.048}^{+0.048}$ & 0.001$_{-4.137}^{+0.000}$ & 0.998$_{-0.005}^{+0.001}$ & 3.751$_{-0.073}^{+0.081}$ & 0.576$_{-0.073}^{+0.090}$ & 0.027$_{-0.000}^{+0.000}$ \\
ZTF20acpevli & 59150.8$_{-0.2}^{+0.9}$ & 12.10$_{-0.089}^{+0.196}$ & 1.063$_{-0.016}^{+0.055}$ & 0.009$_{-0.004}^{+0.000}$ & 0.933$_{-0.024}^{+0.052}$ & 4.986$_{-0.047}^{+0.012}$ & 0.883$_{-0.188}^{+0.040}$ & 0.023$_{-0.000}^{+0.002}$ \\
ZTF20acpvbbh & 59157.3$_{-0.0}^{+0.0}$ & 12.00$_{-0.000}^{+0.002}$ & 1.999$_{-0.002}^{+0.000}$ & 0.009$_{-1.107}^{+3.896}$ & 0.999$_{-0.000}^{+6.792}$ & 3.750$_{-0.001}^{+0.002}$ & 0.085$_{-0.017}^{+0.022}$ & 0.033$_{-0.000}^{+0.000}$ \\
ZTF20acqexmr & 59161.8$_{-0.1}^{+0.1}$ & 12.03$_{-0.033}^{+0.109}$ & 1.514$_{-0.013}^{+0.014}$ & 0.001$_{-0.000}^{+0.000}$ & 0.996$_{-0.012}^{+0.003}$ & 3.785$_{-0.513}^{+0.578}$ & 2.409$_{-0.100}^{+0.097}$ & 0.011$_{-0.000}^{+0.000}$ \\
ZTF20acrinvz & 59166.5$_{-0.4}^{+0.4}$ & 12.80$_{-0.482}^{+0.447}$ & 1.047$_{-0.058}^{+0.071}$ & 0.000$_{-0.000}^{+0.000}$ & 0.914$_{-0.328}^{+0.072}$ & 3.507$_{-1.230}^{+0.750}$ & 0.016$_{-0.015}^{+0.065}$ & 0.041$_{-0.001}^{+0.001}$ \\
ZTF20actawpa & 59169.1$_{-0.0}^{+0.0}$ & 12.00$_{-0.002}^{+0.007}$ & 1.926$_{-0.013}^{+0.012}$ & 0.009$_{-5.775}^{+1.498}$ & 0.999$_{-0.000}^{+0.000}$ & 3.750$_{-0.007}^{+0.008}$ & 0.349$_{-0.015}^{+0.015}$ & 0.022$_{-0.000}^{+0.000}$ \\
ZTF20actnuls & 59169.1$_{-0.3}^{+0.2}$ & 12.40$_{-0.346}^{+0.383}$ & 1.677$_{-0.040}^{+0.062}$ & 0.009$_{-0.001}^{+0.000}$ & 0.998$_{-0.004}^{+0.001}$ & 3.810$_{-0.217}^{+1.094}$ & 1.245$_{-0.074}^{+0.075}$ & 0.026$_{-0.001}^{+0.001}$ \\
ZTF20actpavu & 59169.3$_{-0.9}^{+1.1}$ & 14.21$_{-1.537}^{+1.102}$ & 1.068$_{-0.366}^{+0.863}$ & 0.003$_{-0.001}^{+0.004}$ & 0.922$_{-0.198}^{+0.063}$ & 4.499$_{-1.059}^{+0.427}$ & 0.651$_{-0.305}^{+0.185}$ & 0.027$_{-0.003}^{+0.006}$ \\
ZTF20actquzl & 59170.8$_{-0.1}^{+0.1}$ & 12.01$_{-0.010}^{+0.036}$ & 1.796$_{-0.057}^{+0.059}$ & 0.009$_{-0.000}^{+5.665}$ & 0.998$_{-0.004}^{+0.001}$ & 4.991$_{-0.028}^{+0.008}$ & 0.418$_{-0.043}^{+0.051}$ & 0.033$_{-0.000}^{+0.000}$ \\
ZTF20acuhren & 59175.8$_{-0.5}^{+0.4}$ & 12.99$_{-0.338}^{+0.382}$ & 0.880$_{-0.081}^{+0.161}$ & 0.009$_{-0.004}^{+0.000}$ & 0.423$_{-0.128}^{+0.092}$ & 1.750$_{-0.276}^{+0.747}$ & 0.929$_{-0.078}^{+0.075}$ & 0.019$_{-0.000}^{+0.001}$ \\
ZTF20acvjagm & 59176.2$_{-0.0}^{+0.0}$ & 15.66$_{-0.095}^{+0.094}$ & 1.064$_{-0.021}^{+0.012}$ & 0.009$_{-0.000}^{+4.589}$ & 0.999$_{-0.002}^{+0.000}$ & 4.951$_{-0.036}^{+0.024}$ & 0.343$_{-0.038}^{+0.019}$ & 0.020$_{-0.000}^{+0.000}$ \\
ZTF20acxtdcm & 59191.2$_{-0.6}^{+0.9}$ & 13.45$_{-0.455}^{+0.489}$ & 1.226$_{-0.081}^{+0.102}$ & 0.007$_{-0.002}^{+0.001}$ & 0.974$_{-0.036}^{+0.021}$ & 4.940$_{-0.317}^{+0.054}$ & 0.656$_{-0.201}^{+0.160}$ & 0.040$_{-0.002}^{+0.002}$ \\
ZTF20acyqzeu & 59195.8$_{-0.0}^{+0.0}$ & 12.05$_{-0.045}^{+0.060}$ & 1.228$_{-0.011}^{+0.012}$ & 0.009$_{-1.626}^{+4.628}$ & 0.999$_{-0.000}^{+0.000}$ & 4.999$_{-0.001}^{+0.000}$ & 0.867$_{-0.013}^{+0.013}$ & 0.016$_{-0.000}^{+0.000}$ \\
ZTF21aaabwem & 59201.6$_{-0.3}^{+0.9}$ & 13.33$_{-1.124}^{+0.540}$ & 1.964$_{-0.077}^{+0.032}$ & 0.009$_{-0.001}^{+0.000}$ & 0.995$_{-0.096}^{+0.004}$ & 4.912$_{-0.291}^{+0.081}$ & 0.674$_{-0.178}^{+0.133}$ & 0.049$_{-0.002}^{+0.002}$ \\
ZTF21aabfwwl & 59211.6$_{-0.4}^{+0.2}$ & 13.64$_{-0.172}^{+0.184}$ & 1.002$_{-0.002}^{+0.007}$ & 0.000$_{-7.377}^{+0.000}$ & 0.999$_{-0.001}^{+0.000}$ & 4.992$_{-0.024}^{+0.006}$ & 0.738$_{-0.110}^{+0.067}$ & 0.026$_{-0.000}^{+0.000}$ \\
ZTF21aafepon & 59223.5$_{-0.0}^{+0.0}$ & 14.00$_{-0.037}^{+0.053}$ & 1.262$_{-0.270}^{+0.587}$ & 0.009$_{-0.000}^{+0.000}$ & 0.499$_{-0.003}^{+0.003}$ & 3.750$_{-0.015}^{+0.013}$ & 1.483$_{-0.043}^{+0.046}$ & 0.009$_{-0.000}^{+0.000}$ \\
ZTF21aafkktu & 59225.9$_{-0.2}^{+0.2}$ & 13.15$_{-0.308}^{+0.322}$ & 1.083$_{-0.033}^{+0.031}$ & 0.003$_{-0.000}^{+0.000}$ & 0.995$_{-0.011}^{+0.003}$ & 4.675$_{-0.423}^{+0.211}$ & 0.011$_{-0.010}^{+0.043}$ & 0.037$_{-0.000}^{+0.000}$ \\
ZTF21aafkwtk & 59225.5$_{-2.0}^{+0.1}$ & 12.81$_{-0.252}^{+3.101}$ & 1.363$_{-0.286}^{+0.037}$ & 0.002$_{-0.000}^{+0.000}$ & 0.999$_{-0.008}^{+0.000}$ & 3.779$_{-0.101}^{+1.198}$ & 0.592$_{-0.100}^{+0.281}$ & 0.028$_{-0.002}^{+0.000}$ \\
ZTF21aagbeah & 59225.8$_{-0.0}^{+0.1}$ & 12.07$_{-0.067}^{+0.193}$ & 1.989$_{-0.031}^{+0.009}$ & 0.009$_{-0.000}^{+0.000}$ & 0.997$_{-0.008}^{+0.002}$ & 2.498$_{-0.130}^{+0.133}$ & 0.089$_{-0.051}^{+0.093}$ & 0.058$_{-0.002}^{+0.002}$ \\
ZTF21aagnzjy & 59230.7$_{-0.4}^{+0.3}$ & 12.00$_{-0.004}^{+0.017}$ & 1.502$_{-0.020}^{+0.010}$ & 0.004$_{-0.000}^{+0.001}$ & 0.876$_{-0.043}^{+0.028}$ & 4.122$_{-0.585}^{+0.299}$ & 1.058$_{-0.029}^{+0.046}$ & 0.013$_{-0.000}^{+0.000}$ \\
ZTF21aagsysd & 59233.0$_{-0.1}^{+0.1}$ & 12.00$_{-0.002}^{+0.008}$ & 1.999$_{-0.001}^{+0.000}$ & 0.009$_{-2.267}^{+5.476}$ & 0.999$_{-0.000}^{+0.000}$ & 3.750$_{-0.002}^{+0.002}$ & 1.052$_{-0.029}^{+0.031}$ & 0.023$_{-0.000}^{+0.000}$ \\
ZTF21aagtekf & 59237.1$_{-2.4}^{+1.1}$ & 12.01$_{-0.016}^{+0.054}$ & 1.991$_{-0.027}^{+0.008}$ & 0.009$_{-0.000}^{+0.000}$ & 0.791$_{-0.064}^{+0.068}$ & 4.985$_{-0.054}^{+0.013}$ & 0.023$_{-0.018}^{+0.093}$ & 0.068$_{-0.002}^{+0.001}$ \\
ZTF21aaiaeri & 59245.2$_{-0.1}^{+0.4}$ & 12.01$_{-0.012}^{+0.052}$ & 1.995$_{-0.015}^{+0.004}$ & 0.008$_{-0.000}^{+0.000}$ & 0.999$_{-0.003}^{+0.000}$ & 3.734$_{-1.153}^{+0.122}$ & 0.407$_{-0.078}^{+0.088}$ & 0.047$_{-0.001}^{+0.001}$ \\
ZTF21aaiaqhh & 59246.4$_{-0.0}^{+0.0}$ & 12.00$_{-0.000}^{+0.002}$ & 1.999$_{-0.001}^{+0.000}$ & 0.009$_{-1.055}^{+3.539}$ & 0.999$_{-0.000}^{+3.372}$ & 3.750$_{-0.001}^{+0.002}$ & 0.040$_{-0.022}^{+0.019}$ & 0.031$_{-0.000}^{+0.000}$ \\
ZTF21aapdulz & 59275.3$_{-0.2}^{+0.2}$ & 12.06$_{-0.059}^{+0.170}$ & 0.513$_{-0.012}^{+0.033}$ & 0.009$_{-0.000}^{+8.187}$ & 0.998$_{-0.004}^{+0.001}$ & 2.470$_{-0.765}^{+0.106}$ & 0.922$_{-0.054}^{+0.051}$ & 0.023$_{-0.000}^{+0.000}$ \\
ZTF21aapegtd & 59276.8$_{-0.1}^{+0.6}$ & 12.00$_{-0.005}^{+0.019}$ & 1.994$_{-0.019}^{+0.004}$ & 0.009$_{-0.000}^{+3.177}$ & 0.999$_{-0.001}^{+0.000}$ & 3.746$_{-1.255}^{+0.031}$ & 0.456$_{-0.096}^{+0.079}$ & 0.040$_{-0.001}^{+0.001}$ \\
\enddata

\end{deluxetable}

\bibliography{sample631}{}
\bibliographystyle{aasjournal}



\end{document}